\pdfoutput=1

\documentclass[iop]{emulateapj}

\slugcomment{}

\shorttitle{Star Formation Activity in the molecular cloud G35.20$-$0.74}
\shortauthors{L.~K. Dewangan}
\begin{document}

\title{Star Formation Activity in the molecular cloud G35.20$-$0.74: onset of cloud-cloud collision}
\author{L.~K. Dewangan\altaffilmark{1}}
\email{lokeshd@prl.res.in}
\altaffiltext{1}{Physical Research Laboratory, Navrangpura, Ahmedabad - 380 009, India.}
\begin{abstract}
To probe the star-formation (SF) processes, we present results of an analysis of the molecular cloud G35.20$-$0.74 (hereafter MCG35.2) 
using multi-frequency observations. 
The MCG35.2 is depicted in a velocity range of 30--40 km s$^{-1}$. An almost horseshoe-like structure embedded within the MCG35.2 is evident 
in the infrared and millimeter images and harbors the previously known sites, ultra-compact/hyper-compact G35.20$-$0.74N H\,{\sc ii} region, Ap2-1, and Mercer 14 at its base. 
The site, Ap2-1 is found to be excited by a radio spectral type of B0.5V star where the distribution of 20 cm and H$\alpha$ emission is surrounded by the extended molecular hydrogen emission. 
Using the {\it Herschel} 160--500 $\mu$m and photometric 1--24 $\mu$m data analysis, several embedded clumps and clusters of young stellar objects (YSOs) are investigated 
within the MCG35.2, revealing the SF activities. 
Majority of the YSOs clusters and massive clumps (500--4250 M$_{\odot}$) are seen toward the horseshoe-like structure. 
The position-velocity analysis of $^{13}$CO emission shows a blue-shifted peak (at 33 km s$^{-1}$) and a red-shifted 
peak (at 37 km s$^{-1}$) interconnected by lower intensity intermediated velocity emission, tracing a broad bridge feature. 
The presence of such broad bridge feature suggests the onset of a collision between molecular components in the MCG35.2.  
A noticeable change in the H-band starlight mean polarization angles has also been observed in the MCG35.2, probably tracing the interaction between molecular components. Taken together, it seems that the cloud-cloud collision process has influenced the birth of massive stars and YSOs clusters in the MCG35.2.
 \end{abstract}
\section{Introduction}
\label{sec:intro}
The birth mechanisms of massive stars ($\gtrsim$ 8 M$_{\odot}$) and clusters of young stars are still being 
debated \citep{zin07,tan14}. 
Using the effort of various Galactic plane surveys spanning from near-infrared (NIR) to radio regime, it has been observed that 
molecular clouds often host dark clouds, infrared shells or bubbles associated with H\,{\sc ii} regions, and 
young star clusters. 
Hence, observationally, the understanding of physical processes in such sites is very challenging, indicating 
the onset of numerous complex phenomena involved in star formation.
At the same time, these sites are extremely promising to probe important observational evidences which can directly allow us to validate the 
existing theoretical scenarios concerning the formation of massive stars and stellar clusters. 
More details about the theoretical mechanisms of massive star formation (MSF) can be found in \citet{zin07} and \citet{tan14}. 
Interestingly, the collisions between molecular clouds can also provide the appropriate conditions needed for MSF and birth of star clusters \citep[e.g.,][]{habe92,furukawa09,ohama10,inoue13,fukui14,fukui16,torii15,torii16,haworth15a,haworth15b}. 
However, the observational evidences of star formation (including massive stars) through the cloud-cloud collisions are still very scarce. 
Therefore, it demands a systematic investigation of a given molecular cloud using the observational data having both large field-of-view and 
high resolution at wavelengths which probe obscuring molecular materials.

Situated at a distance of 2.0 kpc \citep{birks06,zhang09,paron10}, 
G35.20$-$0.74 \citep[V$_{lsr}$=33 km s$^{-1}$;][]{seta98,paron10} is a nearby star-forming complex 
that contains at least two active star-forming 
condensations, G35.20$-$0.74N (hereafter G35.20N) and G35.20$-$0.74S (hereafter G35.20S) 
\citep[also see Figure~4 in][]{mooney95}. 
These condensations are active sites of star formation and contain some well known sources, 
such as EGO G35.20$-$0.74 \citep{cyganowski08}, Ap2-1 \citep{paron10}, and Mercer 14 \citep{froebrich11}. 
In the complex, the G35.20N is very well studied site using various data-sets (including Atacama Large Millimeter/submillimeter Array (ALMA)) and is also considered as main site of MSF activity \citep[see][and references therein]{sanchez14}. In the G35.20N, \citet{qiu13} found the presence of a parsec-sized and wide-angle molecular outflow driven from a strongly rotating region 
whose magnetic field is largely toroidal. 
Using the ALMA 870 $\mu$m data (350 GHz; resolution $\sim$$0\farcs$4), an elongated dust structure in the G35.20N was investigated and 
has been found to be fragmented into a number of dense cores \citep{sanchez14}. They further suggested that these dense cores might form 
massive stars. Using the Very Large Array (VLA) and ALMA observational data-sets, \citet{beltran16} also investigated a binary system of 
ultra-compact/hyper-compact H\,{\sc ii} regions at the geometrical center of the radio jet in G35.20N. 
\citet{paron10} found that Ap2-1 is an H\,{\sc ii} region excited by an early B-type star and is associated with 
the extended 8.0 $\mu$m emission. They further reported a radial velocity of V$_{lsr}$ $\sim$31 km s$^{-1}$ for Ap2-1 using the H$\alpha$ and [N{\sc ii}] lines detected in its optical low-resolution spectrum \citep[see Figure~9 in][]{paron10}. \citet{froebrich11} reported several molecular hydrogen outflows in Mercer 14 and suggested 
ongoing star formation in this site \citep[see Figure~1 in][]{froebrich11}. 
Several Spitzer dark clouds were reported in the G35.20$-$0.74 complex \citep{peretto09}. 
Together, these previous studies indicate the ongoing star formation activity in the G35.20$-$0.74 region as well as 
the formation of young massive stars. Note that the previous works were mainly concentrated toward 
the sites, G35.20N and G35.20S (including Ap2-1 and Mercer 14). 

The study of star formation processes in {\it the entire molecular cloud associated with the star-forming complex, G35.20$-$0.74 
(hereafter MCG35.2)} is yet to be probed. 
Despite the presence of various observational investigations, 
the physical processes responsible for star formation in the MCG35.2 are not well explored.
A careful investigation of velocity structure of molecular gas in the MCG35.2 is still missing in the literature. 
The large scale magnetic field structure projected in the sky plane toward the MCG35.2 is yet to be examined. 
The objective of this paper is to study the physical conditions in the MCG35.2 and to infer the physical mechanisms operating within the molecular cloud.
In this paper, we present the results of the MCG35.2 extracted using a more detailed and extensive analysis of observational 
data from centimeter, NIR, and optical H$\alpha$ wavelengths. 
These multi-frequency data-sets are retrieved from various Galactic plane surveys (see Table~\ref{ftab1}). 
Using these multi-frequency data-sets, we probe the distribution of dust temperature, column density, extinction, ionized emission, 
molecular hydrogen emission, magnetic field lines, kinematics of $^{13}$CO gas, 
and young stellar objects (YSOs) in the MCG35.2.  

This paper is organized as follows. In Section~\ref{sec:obser}, we describe the data selection. 
In Section~\ref{sec:data}, we present the outcomes of our extensive multi-frequency data analysis. 
In Section~\ref{sec:disc}, we discuss the implications of our results concerning the formation of massive stars and young stellar clusters. 
Finally, the findings are summarized and concluded in Section~\ref{sec:conc}.
\section{Data and analysis}
\label{sec:obser}
In this work, we have chosen a region of $\sim$42$\farcm5$ $\times$ 30$\farcm3$ 
(central coordinates: $l$ = 35$\degr$.144; $b$ = $-$0$\degr$.845) around the MCG35.2.
In the following, we give a brief description of the adopted multi-frequency data-sets (also see Table~\ref{ftab1}). 
\subsection{Radio Centimeter Continuum Maps}
We used radio continuum maps at MAGPIS 20 cm and NVSS 1.4 GHz.
The MAGPIS 20 cm map has a 6\farcs2 $\times$ 5\farcs4 beam size and a pixel scale of 2$\arcsec$/pixel. 
The beam size of the NVSS 1.4 GHz map is $\sim$45$\arcsec$. 
These maps were obtained with a typical rms of 0.3 -- 0.45 mJy/beam. 
The centimeter radio continuum map traces the ionized emission in the complex.
\subsection{$^{13}$CO (J=1$-$0) Line Data}
In order to trace molecular gas associated with the selected target, 
the GRS $^{13}$CO (J=1$-$0) line data were employed. 
The GRS line data have a velocity resolution of 0.21~km\,s$^{-1}$, an angular resolution 
of 45$\arcsec$ with 22$\arcsec$ sampling, a main beam efficiency ($\eta_{\rm mb}$) of $\sim$0.48, 
a velocity coverage of $-$5 to 135~km~s$^{-1}$, and a typical rms sensitivity (1$\sigma$)
of $\approx0.13$~K \citep{jackson06}.  
\subsection{{\it Herschel} and ATLASGAL Data}
To infer the far-infrared (FIR) and sub-millimeter (mm) emission, 
{\it Herschel} continuum maps at 70 $\mu$m, 160 $\mu$m, 250 $\mu$m, 350 $\mu$m, and 500 $\mu$m were retrieved 
from the {\it Herschel} data archive. The beam sizes of these images are 5$\farcs$8, 12$\arcsec$, 18$\arcsec$, 25$\arcsec$, and 37$\arcsec$ \citep{poglitsch10,griffin10}, respectively. The processed level2$_{-}$5 products were selected using the {\it Herschel} Interactive Processing 
Environment \citep[HIPE,][]{ott10}. 

The sub-mm continuum map at 870 $\mu$m (beam size $\sim$19$\farcs$2) was also obtained from the ATLASGAL. 
\subsection{{\it Spitzer} and {\it WISE} Data}
The photometric images and magnitudes of point sources at 3.6--8.0 $\mu$m were obtained 
from the {\it Spitzer}-GLIMPSE survey (resolution $\sim$2$\arcsec$). 
In this work, we downloaded the GLIMPSE-I Spring '07 highly reliable photometric catalog. 
We also used MIPSGAL 24 $\mu$m image and the photometric magnitudes of point sources at MIPSGAL 24 $\mu$m \citep[from][]{gutermuth15}. 

We also utilized the publicly available archival WISE\footnote[1]{WISE is a joint project of the
University of California and the JPL, Caltech, funded by the NASA} image at mid-infrared (MIR) 12 $\mu$m 
(spatial resolution $\sim$6$\arcsec$). 
\subsection{H$_{2}$ Narrow-band Image}
Continuum-subtracted narrow-band H$_{2}$ (v = $1-0$ S(1)) image at 2.12 $\mu$m 
was obtained from the UWISH2 database. 
The UWISH2 is an unbiased survey of the inner Galactic plane in the H$_{2}$ 1-0 S(1) line using 
the UKIRT Wide Field Camera \citep[WFCAM;][]{casali07}. 
The v= 1-0 S(1) line of H$_{2}$ at 2.12 $\mu$m is an excellent tracer of shocked regions. 
\subsection{NIR Data}
We retrieved NIR photometric {\it JHK} magnitudes of point sources from the UKIDSS GPS 6$^{th}$ archival data release (UKIDSSDR6plus). 
The UKIDSS observations (resolution $\sim$$0\farcs8$) were carried out using the WFCAM at UKIRT. 
The final fluxes were calibrated using the 2MASS data. Several bright sources were found to be saturated in the catalog. 
In this work, we obtained only reliable NIR photometric catalog. 
More information about the selection procedure of the GPS photometry can be obtained in \citet{dewangan15}.
Our selected GPS catalog contains point-sources fainter than J = 12.5, H = 11.7, and K = 10.6 mag to avoid saturation. 
2MASS data were adopted for bright sources that were saturated in the GPS catalog. 
\subsection{H-band Polarimetry}
NIR H-band (1.6 $\mu$m) polarimetric data (resolution $\sim$$1\farcs5$) toward the MCG35.2 were obtained from the 
GPIPS. The region probed in this paper was covered in twelve GPIPS fields (i.e., 
GP1407, GP1408, GP1421, GP1422, GP1423, GP1435, GP1436, GP1437, GP1450, GP1451, GP1452, and GP1464). 
The GPIPS data were observed with the {\it Mimir} instrument, mounted on the 1.8 m Perkins telescope, 
in H-band linear imaging polarimetry mode \citep[see][for more details]{clemens12}. 
To obtain the reliable polarimetric results, we selected sources having Usage Flag (UF) = 1 
and $P/\sigma_p \ge$ 2. These conditions yielded a total of 610 stars. 
The polarization vectors of background stars trace the magnetic field direction in the plane of the sky parallel to the direction of
polarization \citep{davis51}.
\subsection{H$\alpha$ Narrow-band Image}
We downloaded narrow-band H$\alpha$ image at 0.6563 $\mu$m from the IPHAS survey database. 
The survey was made with the Wide-Field Camera (WFC) at the 2.5-m INT, located at La Palma. 
The WFC consists of four 4k $\times$ 2k CCDs, in an L-shape configuration. The pixel
scale is $0\farcs33$ (see \citet{drew05} for more details). 
The H$\alpha$ map depicts the distribution of ionized emission. 
\section{Results}
\label{sec:data}
\subsection{Multi-wavelength picture of MCG35.2}
\label{sec:hh}
\begin{figure*}
\epsscale{1}
\plotone{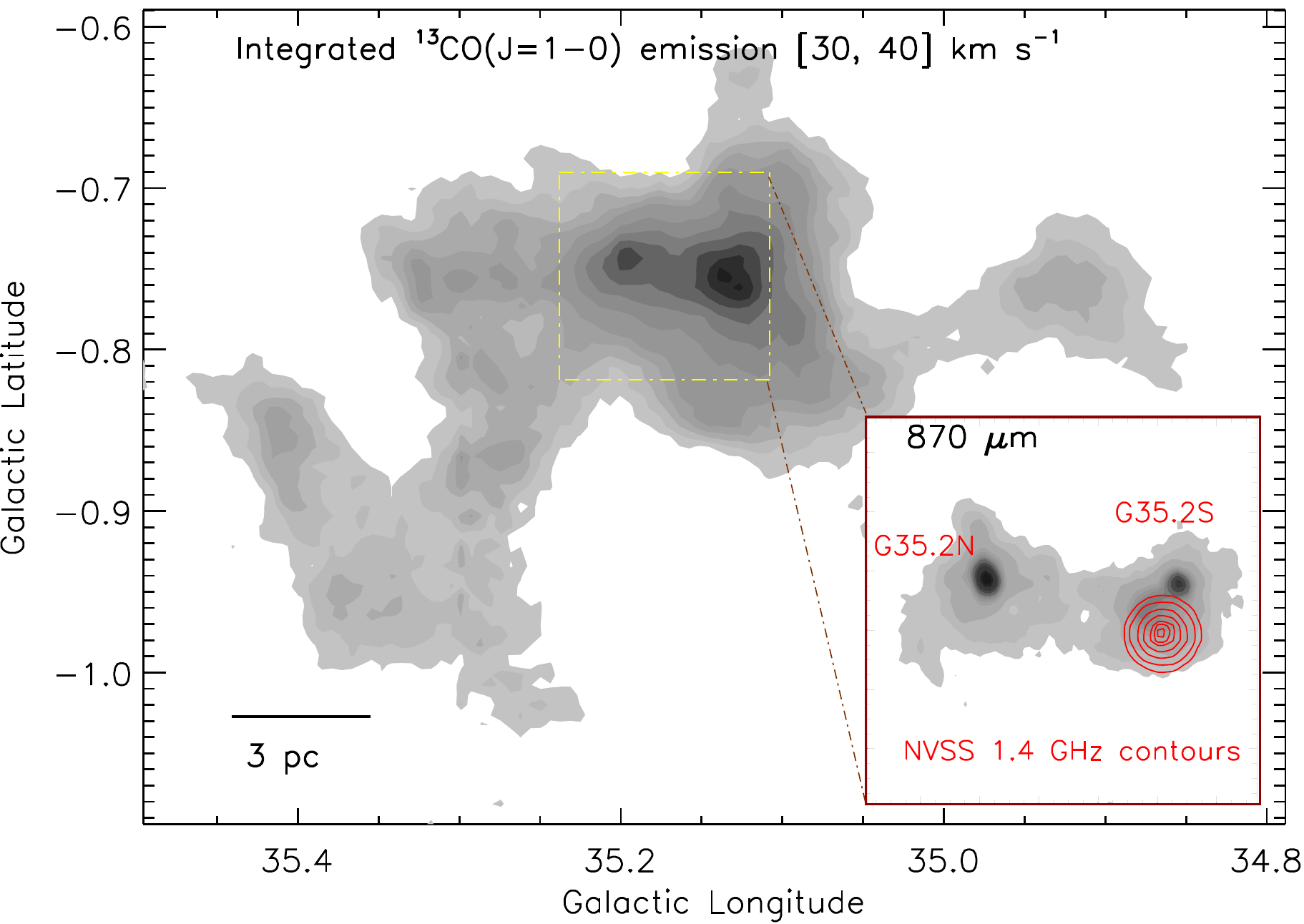}
\caption{\scriptsize Molecular cloud MCG35.2: integrated $^{13}$CO (1-0) emission from 30 to 40 km s$^{-1}$ 
(size of the selected field $\sim$42$\farcm5$ $\times$ 30$\farcm3$ ($\sim$24.7 pc $\times$ 17.6 pc at a distance of 2.0 kpc); 
centered at $l$ = 35$\degr$.144; $b$ = $-$0$\degr$.845). 
The CO contours are 61.78 K km s$^{-1}$ $\times$ (0.2, 0.25, 0.3, 0.35, 0.4, 0.5, 0.6, 0.7, 0.8, 0.9, 0.98).
The inset on the bottom right shows the central region in zoomed-in view, using the ATLASGAL 870 $\mu$m contour map 
overlaid with the NVSS 1.4 GHz contours (see a dotted-dashed yellow box in figure). 
The NVSS contours are superimposed with levels of 10, 20, 40, 60, 80, 90, and 98\% of 
the peak value (i.e., 0.1653 Jy/beam). The locations of molecular condensations G35.2N and G35.2S are also highlighted. 
The scale bar at the bottom-left corner corresponds to 3 pc (at a distance of 2.0 kpc).}
\label{fig1}
\end{figure*}

\begin{figure*}
\epsscale{0.78}
\plotone{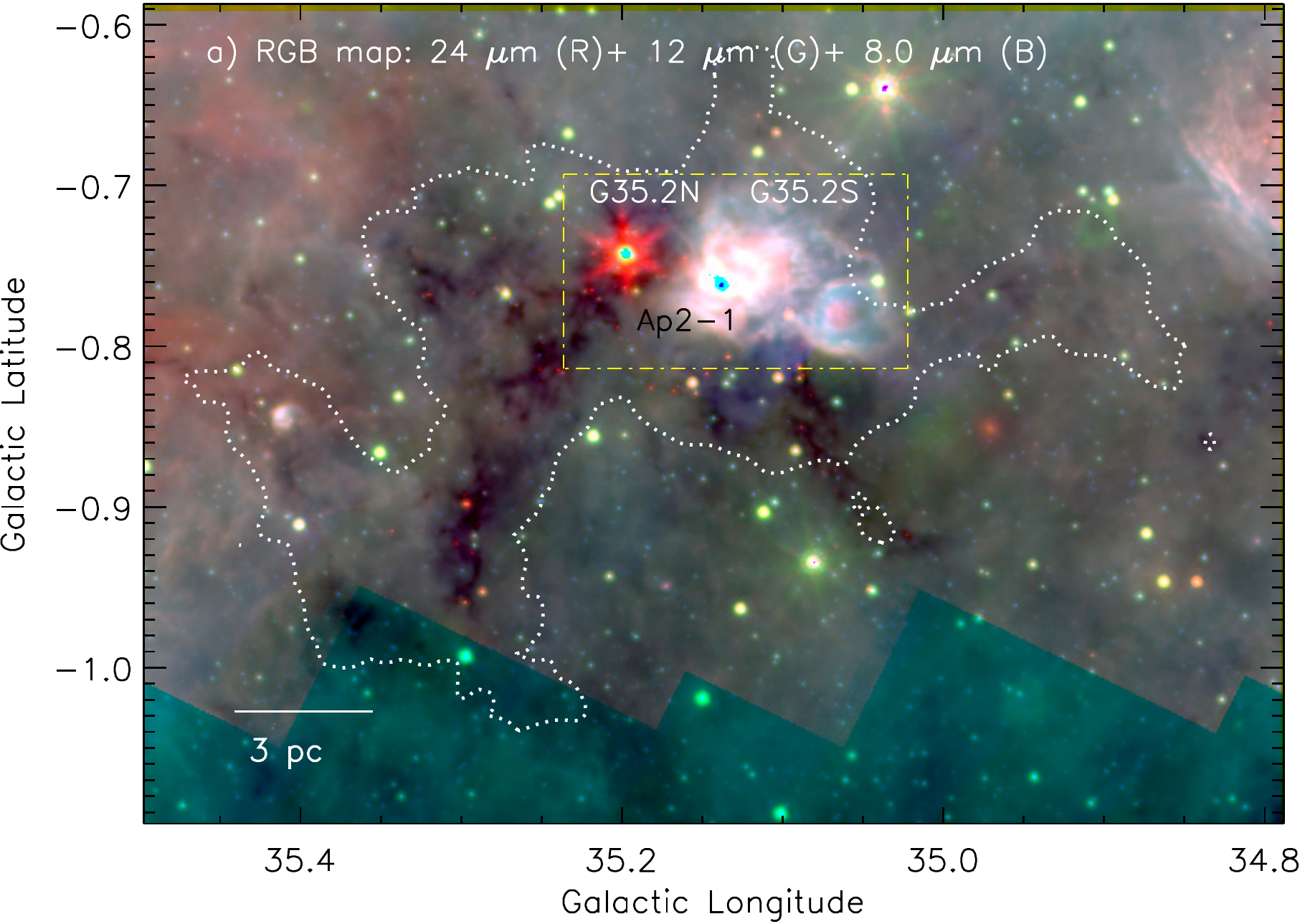}
\epsscale{0.78}
\plotone{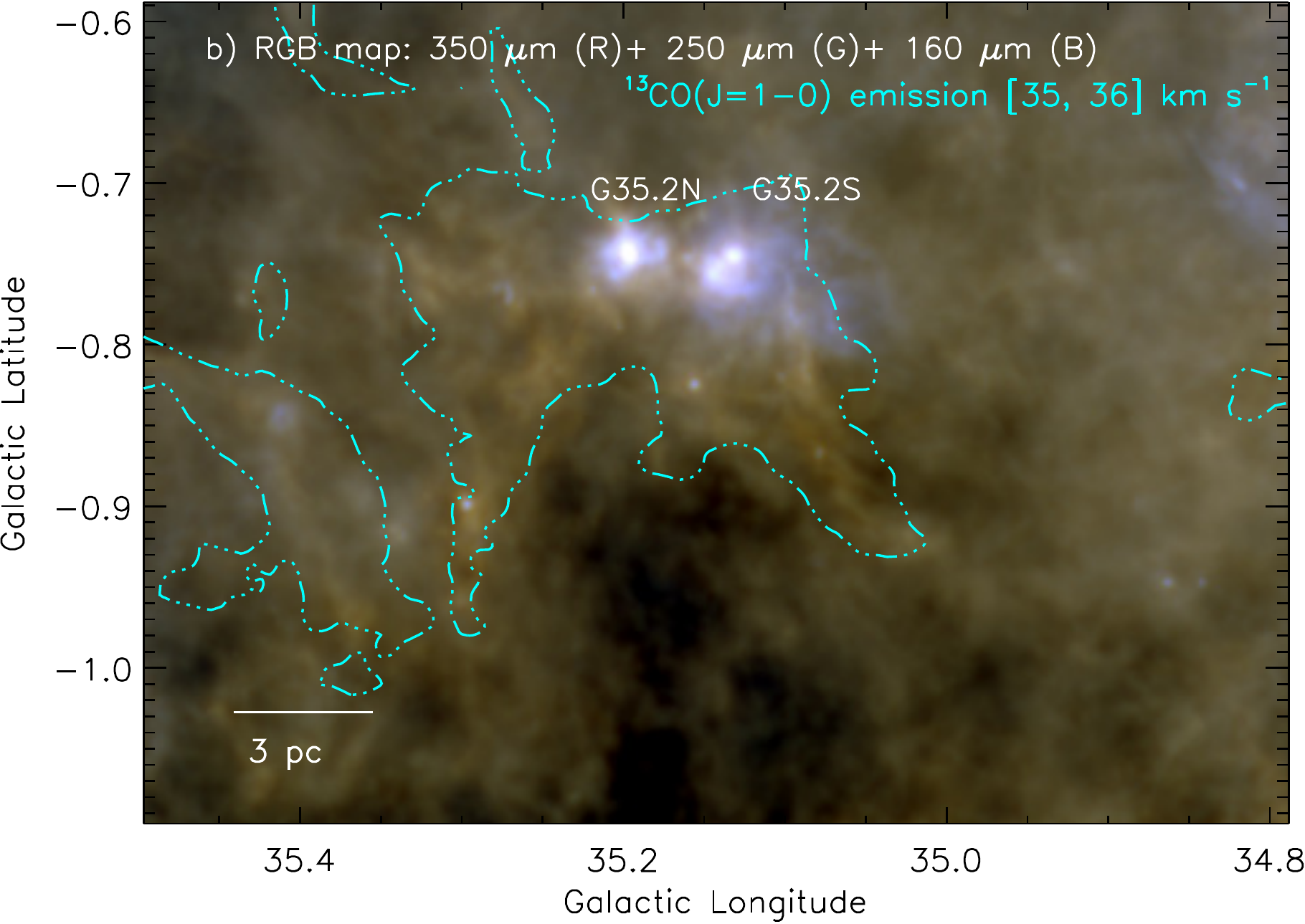}
\caption{\scriptsize Three-color images of the star-forming complex, G35.20$-$0.74 using the {\it Spitzer}, {\it WISE}, and {\it Herschel} images.
a) The image is a result of the combination of three bands: 24 $\mu$m (red), 12 $\mu$m (green), 
and 8.0 $\mu$m (blue). The CO contour (dotted white) is shown with a level of 12.356 K km s$^{-1}$. 
The locations of G35.2N, G35.2S, and Ap2-1 are highlighted.
Figure reveals the dark clouds located within the MCG35.2 and the distribution of dark clouds appears as a horseshoe-like morphology.  
A dotted-dashed box shows the field of Figure~\ref{fig3}a.
b) Color-composite map using the {\it Herschel} 350 $\mu$m (red), 250 $\mu$m (green), and 160 $\mu$m (blue) images. 
The CO emission contour (dotted-dashed cyan) is also shown from 35 to 36 km s$^{-1}$ with a level of 3 K km s$^{-1}$, 
tracing an almost horseshoe-like morphology containing the sources G35.2N and G35.2S at its base (also see Figure~\ref{fig5}). The scale bar corresponding to 3 pc (at a distance of 2.0 kpc) is shown in both the panels.}
\label{fig2}
\end{figure*}

\begin{figure*}
\epsscale{0.73}
\plotone{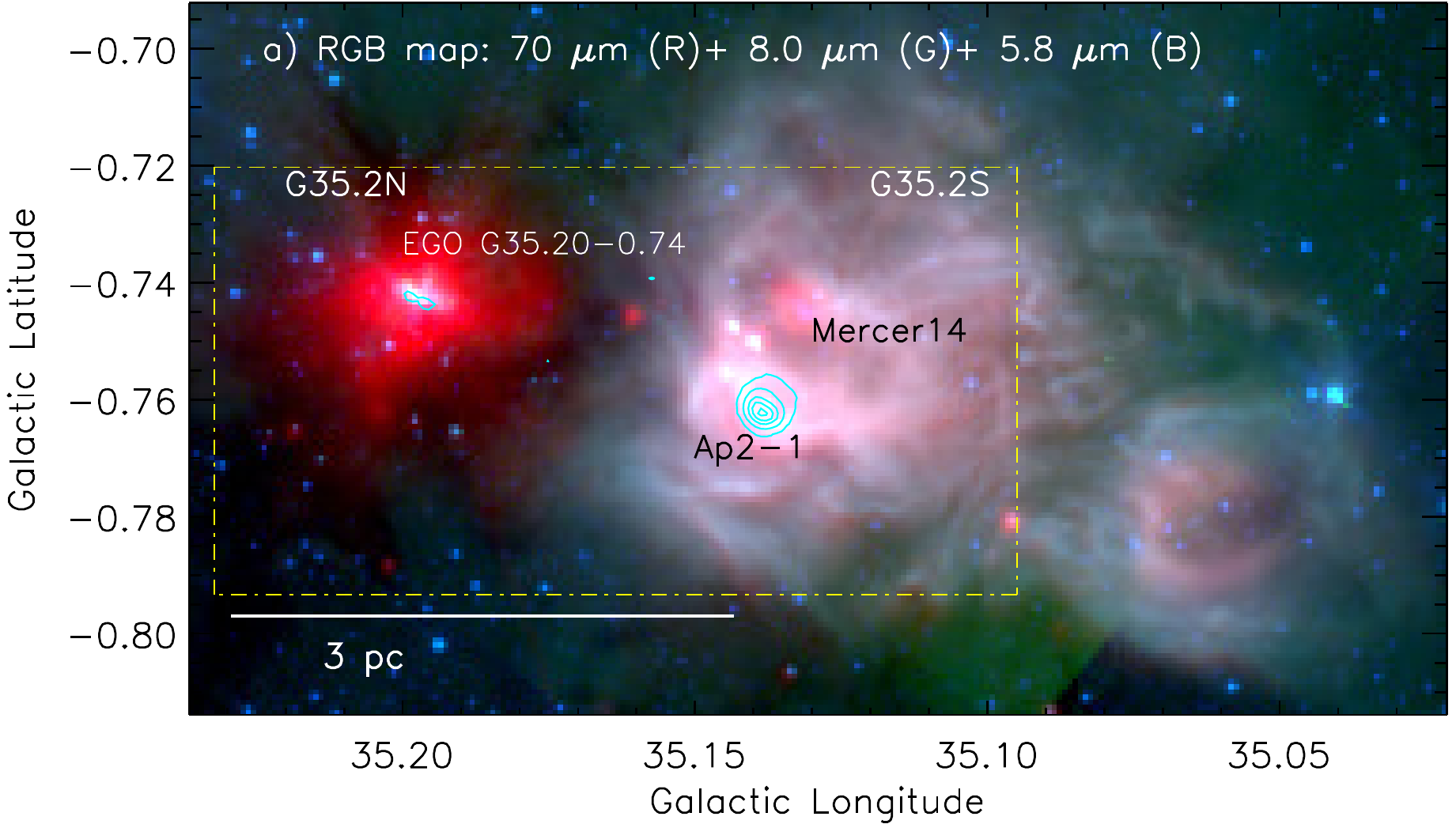}
\epsscale{0.73}
\plotone{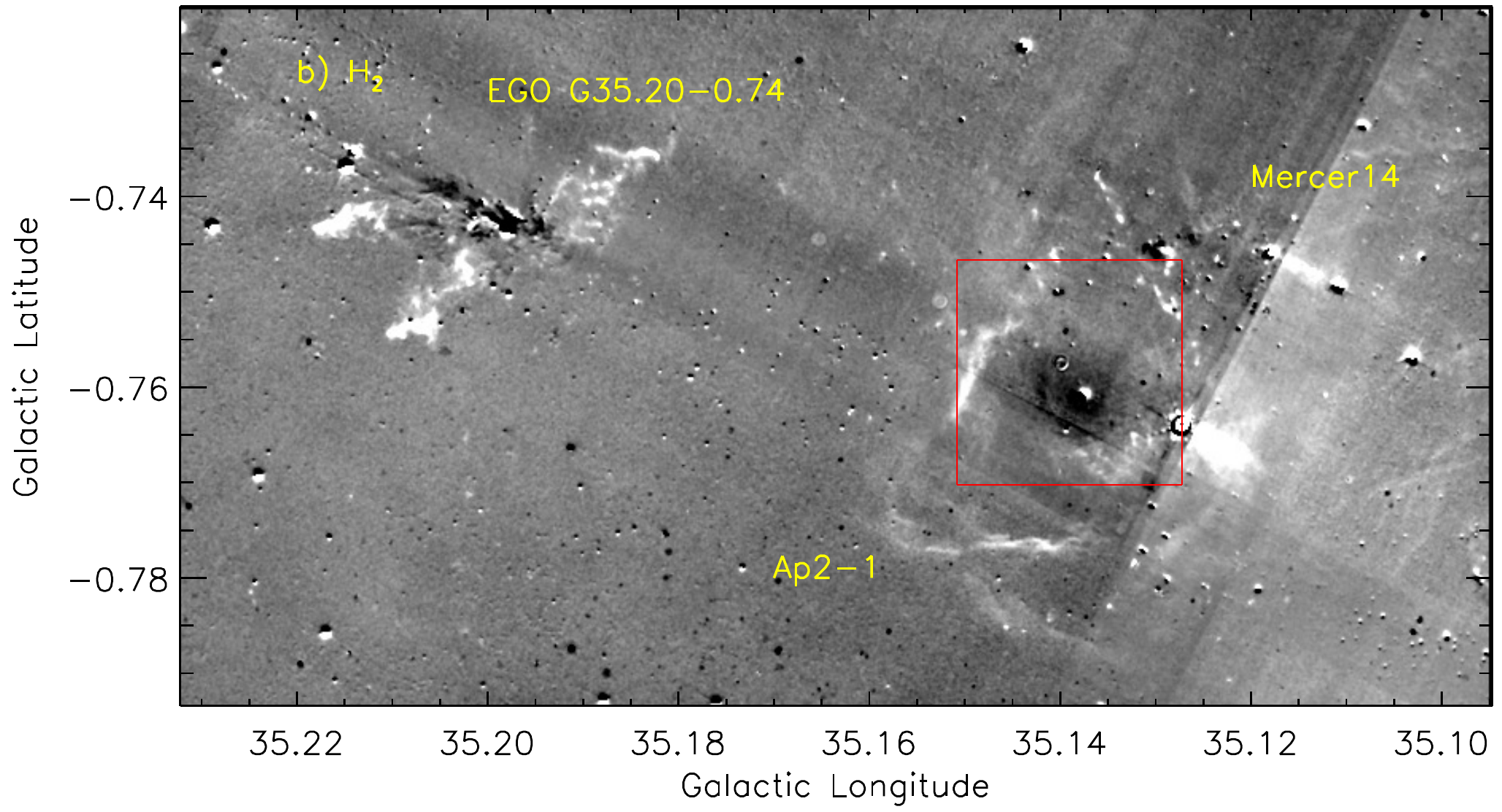}
\epsscale{0.73}
\plotone{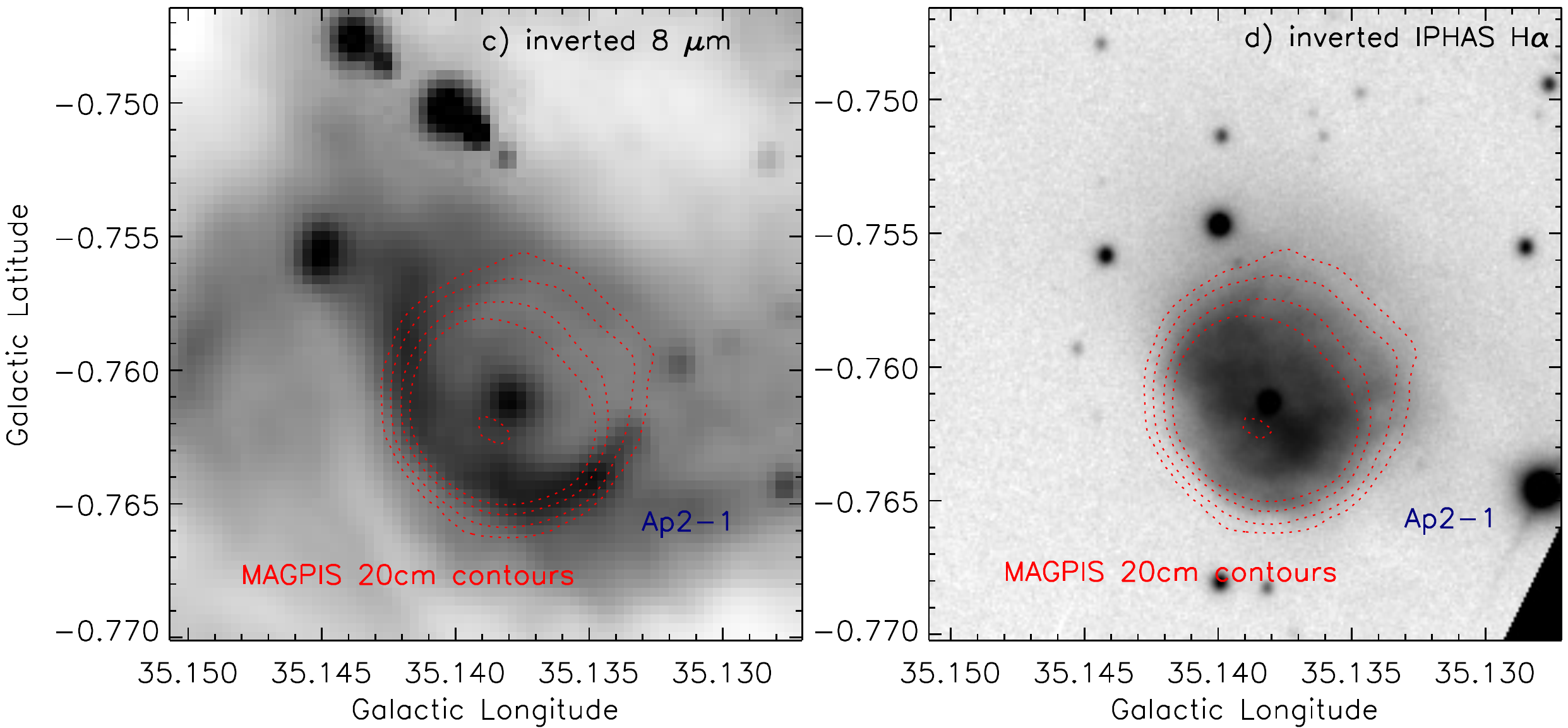}
\caption{\scriptsize Zoomed-in view of G35.2N, G35.2S, and Ap2-1. 
a) The image is a result of the combination of three bands: 70 $\mu$m (red), 8.0 $\mu$m (green), 
and 5.8 $\mu$m (blue). The MAGPIS 20 cm emission is overplotted by cyan solid contours with levels of 5, 30, 60, 80, and 95\% of 
the peak value (i.e., 0.0257 Jy/beam). A dotted-dashed box shows the field of Figure~\ref{fig3}b.
b) The UWISH2 continuum-subtracted H$_{2}$ map (gray-scale) at 2.12 $\mu$m (see dotted-dashed box in Figure~\ref{fig3}a). 
A solid box (in red) shows the field of Figures~\ref{fig3}c and~\ref{fig3}d. 
c) Zoomed-in view of Ap2-1 nebula using the {\it Spitzer} 8.0 $\mu$m image (inverted gray-scale).
d) Zoomed-in view of Ap2-1 nebula using the IPHAS H$\alpha$ image (inverted gray-scale). 
In last two panels (c and d), the images are overlaid with the MAGPIS 20 cm emission. 
The 20 cm contours are shown with levels of 5, 10, 20, 30, and 98\% of 
the peak value (i.e., 0.0257 Jy/beam).}
\label{fig3}
\end{figure*}

An extensive analysis of multi-wavelength data is essential to study the physical environment of a given star-forming complex.
In Figure~\ref{fig1}, we show the molecular $^{13}$CO (J = 1--0) gas emission in the direction 
of the MCG35.2. The $^{13}$CO profile depicts the MCG35.2 in a velocity range of 30--40 km s$^{-1}$. 
The molecular map traces previously known condensations G35.20N and G35.20S that are also seen in the ATLASGAL 870 $\mu$m map.
The NVSS 1.4 GHz emission is detected toward the G35.20S (see Figure~\ref{fig1}). 
Considering the previous results (such as V$_{lsr}$) on G35.20N and Ap2-1 sources \citep{sanchez14,paron10}, we infer that the condensations 
G35.20N and G35.20S are physically associated with the MCG35.2. In Figure~\ref{fig2}a, we present a three-color composite 
image made using the {\it Spitzer} and {\it WISE} images (24 $\mu$m in red, 12 $\mu$m in green, and 8.0 $\mu$m in blue). 
The image traces several dark clouds within the MCG35.2 and the distribution of these dark clouds appears as an almost horse-shoe like structure.
The horseshoe-like structure of the dark clouds is identified based on the absorption 
features against the Galactic background in the 8--24 $\mu$m images. 
This particular structure is more obvious in the $^{13}$CO emission from 35 to 36 km s$^{-1}$ (see Figure~\ref{fig2}b). 
Several embedded stellar sources are also visually seen 
in the dark clouds at 8--24 $\mu$m. Figure~\ref{fig2}b is a three-color composite image made using the 
{\it Herschel} 350 $\mu$m in red, 250 $\mu$m in green, and 160 $\mu$m in blue. 
The $^{13}$CO emission from 35 to 36 km s$^{-1}$ is also superimposed on the color composite image. 
The dark clouds appear as bright emission regions at wavelengths longer than 70 $\mu$m. 
The distribution of $^{13}$CO emission is used to depict column density in the outer regions of dark clouds, tracing the horseshoe-like structure. 
The condensations G35.20N and G35.20S appear to be situated 
at the base of the horseshoe-like structure. Figure~\ref{fig3}a shows the zoomed-in view toward the G35.20N and G35.20S 
(a color composite image: 70 $\mu$m (red), 8.0 $\mu$m (green), and 5.8 $\mu$m (blue) images). 
The MAGPIS 20 cm emission is also overlaid on the composite image, revealing radio continuum 
detections in EGO G35.20$-$0.74 and Ap2-1. 
In Figure~\ref{fig3}b, the molecular hydrogen outflows are found toward the EGO G35.20$-$0.74 and Mercer 14, which 
were previously known in the literature \citep[e.g.][]{froebrich11,sanchez14}.  An arc-like feature is seen in the {\it Spitzer} 3.6--8.0 $\mu$m, 
which is designated as Ap2-1 nebula \citep[e.g.][]{paron10}. The {\it Spitzer} 8.0 $\mu$m image of Ap2-1 nebula is shown in Figure~\ref{fig3}c and is also 
overlaid with the MAGPIS 20 cm emission. 
In Figure~\ref{fig3}d, we present H$\alpha$ image overlaid with the MAGPIS 20 cm emission and find the similarity in the spatial distribution of radio continuum and H$\alpha$ emission. The 20 cm and H$\alpha$ emissions are coincident with the 8.0 $\mu$m feature. 
The 8.0 $\mu$m feature is surrounded by the extended molecular H$_{2}$ emission,  suggesting 
the origin of the H$_{2}$ emission is likely caused by ultraviolet (UV) fluorescence (see Figure~\ref{fig3}b).

In Figure~\ref{fig3}d, the distribution of ionized emission is almost spherical. 
Using the MAGPIS 20 cm map and the ``CLUMPFIND" IDL program \citep{williams94}, the integrated flux 
density and the radius (R$_{HII}$) of the H\,{\sc ii} region are estimated to be 230.3 mJy and 0.17 pc, respectively. 
Using this integrated flux density value, the number of Lyman continuum photon (N$_{uv}$) is estimated 
following the equation given in \citet{matsakis76} \citep[see][for more details]{dewangan16}. 
This analysis is carried out for a distance of 2.0 kpc and for the electron temperature of 10000~K. 
We compute N$_{uv}$ (or logN$_{uv}$) to be $\sim$7.3 $\times$ 10$^{46}$ s$^{-1}$ (46.86) for the H\,{\sc ii} region associated with Ap2-1, 
corresponding to a single ionizing star of spectral type B0.5V-B0V (see Table II in \citet{panagia73}). 
Furthermore, the dynamical age of the H\,{\sc ii} region associated with Ap2-1 is defined below at a given radius R$_{HII}$ \citep[e.g.,][]{dyson80}:
\begin{equation}
t_{dyn} = \left(\frac{4\,R_{s}}{7\,c_{s}}\right) \,\left[\left(\frac{R_{HII}}{R_{s}}\right)^{7/4}- 1\right] 
\end{equation}
where c$_{s}$ is the isothermal sound velocity in the ionized gas (c$_{s}$ = 11 km s$^{-1}$; \citet{bisbas09}), 
R$_{HII}$ is defined earlier, and R$_{s}$ is the radius of the Str\"{o}mgren sphere (= (3 N$_{uv}$/4$\pi n^2_{\rm{0}} \alpha_{B}$)$^{1/3}$, where 
the radiative recombination coefficient $\alpha_{B}$ =  2.6 $\times$ 10$^{-13}$ (10$^{4}$ K/T)$^{0.7}$ cm$^{3}$ s$^{-1}$ \citep{kwan97}, 
N$_{uv}$ is defined earlier, and ``n$_{0}$'' is the initial particle number density of the ambient neutral gas. 
Adopting typical value of n$_{0}$ (as 10$^{3}$(10$^{4}$) cm$^{-3}$), we estimated the dynamical age of the H\,{\sc ii} region 
to be $\sim$4100(32800) yr.
  
Together, Figures~\ref{fig1}--\ref{fig3} help us to infer the embedded structures in the MCG35.2 and its physical association with the cloud.
\subsection{{\it Herschel} temperature and column density maps}
\label{subsec:temp}
The {\it Herschel} temperature and column density maps of MCG35.2 are presented in this section.
These maps are generated in a manner similar to that outlined in \citet{mallick15}.
Here, we also provide a brief step-by-step description of the procedures. 

We produced the {\it Herschel} temperature and column density maps from a  pixel-by-pixel spectral energy distribution (SED) 
fit with a modified blackbody curve to the cold dust emission in the {\it Herschel} 160--500 $\mu$m wavelengths. 
The {\it Herschel} 70 $\mu$m data are not included in the analysis, because the 70 $\mu$m emission is 
dominated by the UV-heated warm dust. The {\it Herschel} 160 $\mu$m image is calibrated in the units of 
Jy pixel$^{-1}$, while the images at 250--500 $\mu$m are calibrated in the surface brightness unit of MJy sr$^{-1}$.
Plate scales are 3.2 arcsec/pixel for the 160 $\mu$m image, 6 arcsec/pixel for the 250 $\mu$m image, 10 arcsec/pixel 
for the 350 $\mu$m image, and 14 arcsec/pixel for the 500 $\mu$m image. 

Prior to the SED fit, we convolved the images at 160--350 $\mu$m to the lowest angular 
resolution of image at 500 $\mu$m ($\sim$37$\arcsec$) and converted these images into the same flux unit (i.e. Jy pixel$^{-1}$). 
Furthermore, the images at 160--350 $\mu$m were regridded  to the pixel size of image at 500 $\mu$m ($\sim$14$\arcsec$). 
These procedures were carried out using the convolution kernels available in the HIPE software. 
Next, we determined the sky background flux level be 0.211, 0.566, 1.023, and 0.636 Jy pixel$^{-1}$ for the 500, 350, 250, and 
160 $\mu$m images (size of the selected region $\sim$4$\farcm$9 $\times$ 6$\farcm$4; 
centered at:  $l$ = 35$\degr$.208; $b$ = $-$1$\degr$.47), respectively. 
To avoid diffuse emission linked with the MCG35.2, the featureless dark field away from the selected target was 
carefully selected for the background estimation. 

\begin{figure*}
\epsscale{0.487}
\plotone{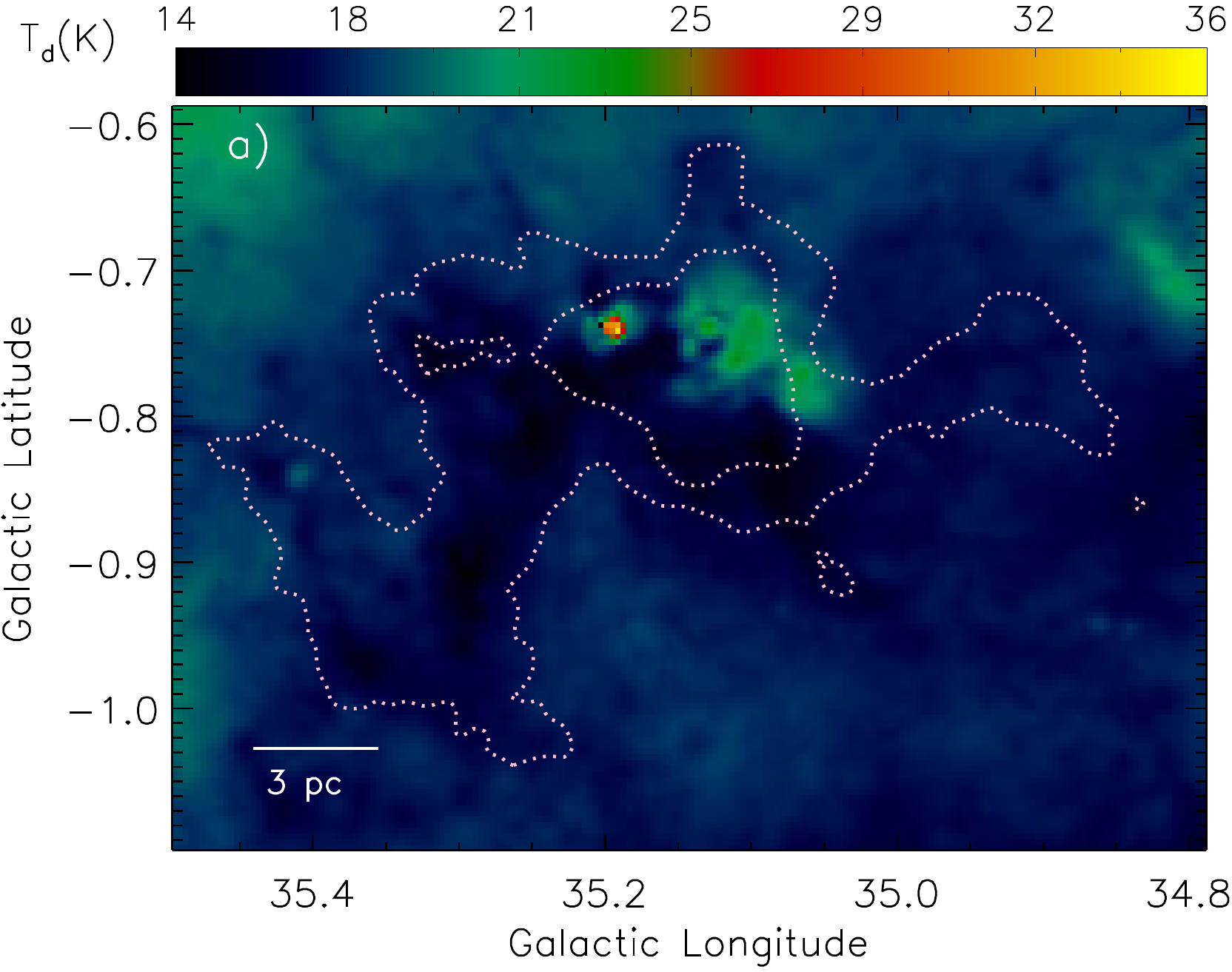}
\epsscale{0.487}
\plotone{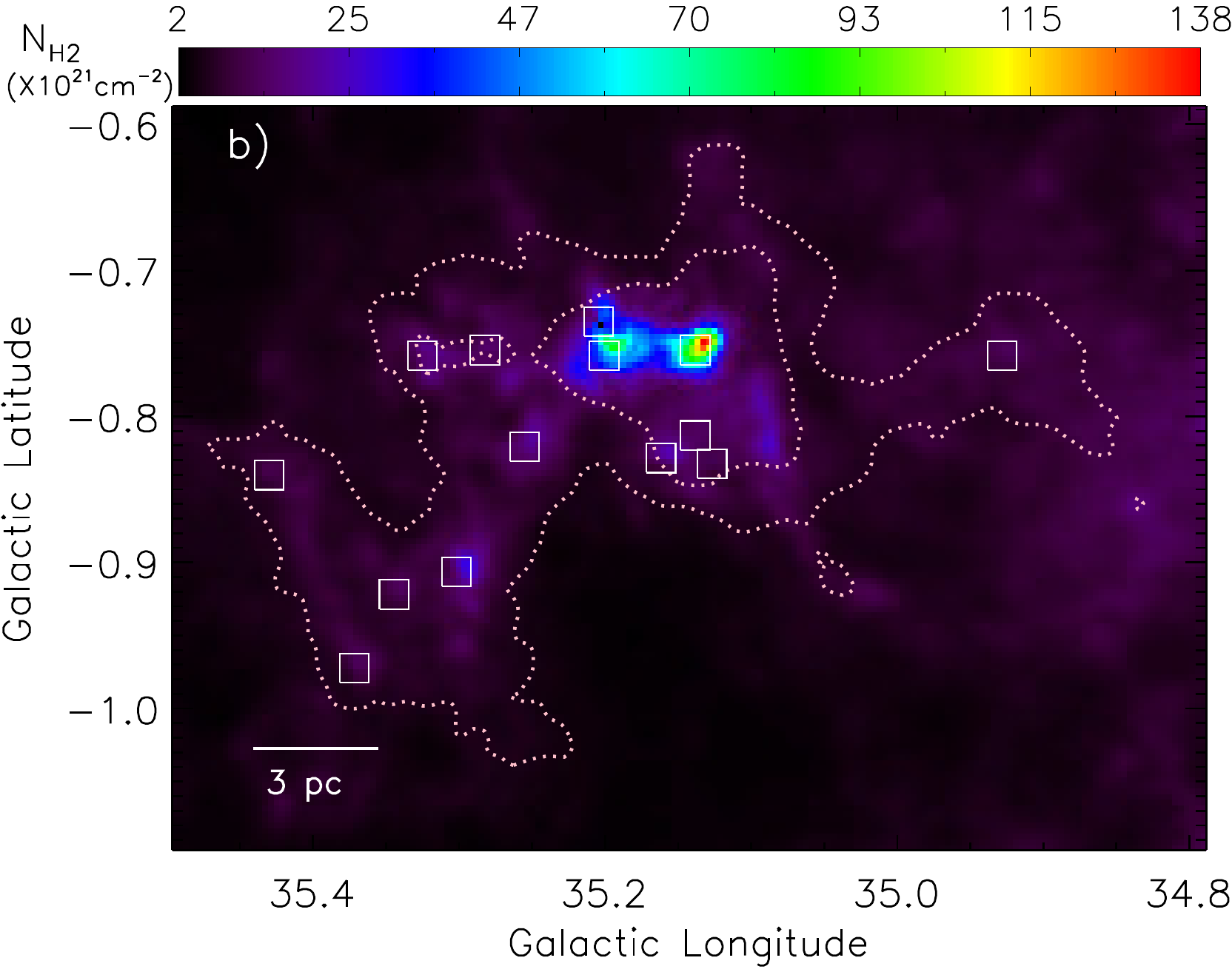}
\epsscale{0.49}
\plotone{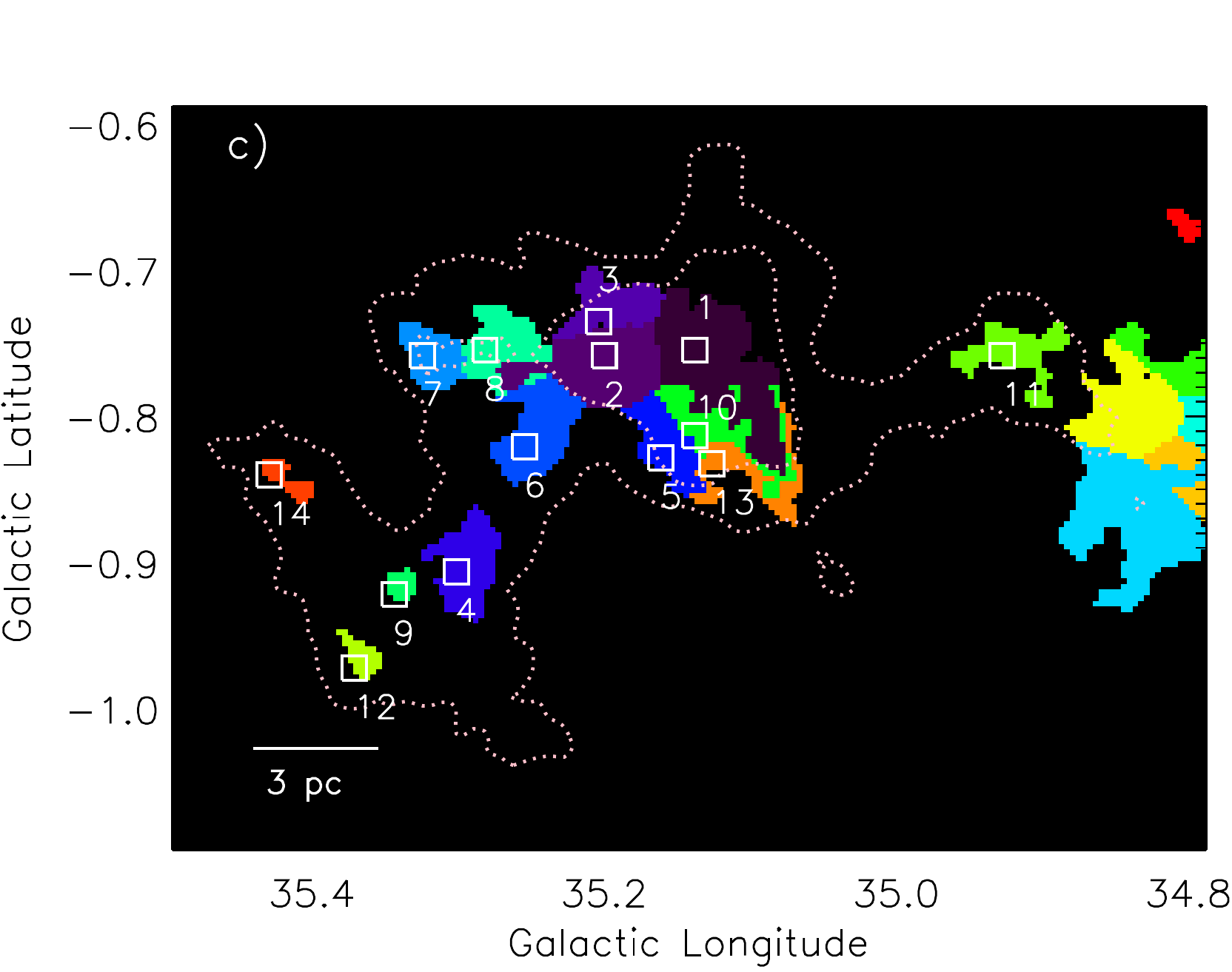}
\caption{\scriptsize {\it Herschel} temperature map (a) and column density ($N(\mathrm H_2)$) map (b) 
of the star-forming complex, G35.20$-$0.74 (see text for details). The column density map permits to identify the clumps and to compute the extinction 
with $A_V=1.07 \times 10^{-21}~N(\mathrm H_2)$ \citep{bohlin78} (see text for details). 
The identified clumps are highlighted by squares and the boundary of each {\it Herschel} clump is 
shown in Figure~\ref{fig4}c. c) The boundary of each {\it Herschel} clump is 
highlighted along with its corresponding clump ID (also see Table~\ref{tab1}). In all the panels, the CO contours (dotted white) are shown (see Figure~\ref{fig1}).
In each panel, the scale bar at the bottom-left corner corresponds to 3 pc (at a distance of 2.0 kpc).}
\label{fig4}
\end{figure*}

Finally, to produce the maps, a modified blackbody was fitted to the observed fluxes on a pixel-by-pixel basis 
\citep[see equations 8 and 9 in][]{mallick15}. 
The fitting was carried out using the four data points for each pixel, maintaining the 
dust temperature (T$_{d}$) and the column density ($N(\mathrm H_2)$) 
as free parameters. 
In the analysis, we adopted a mean molecular weight per hydrogen molecule ($\mu_{H2}$=) 2.8 
\citep{kauffmann08} and an absorption coefficient ($\kappa_\nu$ =) 0.1~$(\nu/1000~{\rm GHz})^{\beta}$ cm$^{2}$ g$^{-1}$, 
including a gas-to-dust ratio ($R_t$ =) of 100, with a dust spectral index of $\beta$\,=\,2 \citep[see][]{hildebrand83}. 
In Figure~\ref{fig4}, the final temperature and column density maps (resolution $\sim$37$\arcsec$) are presented.

In the {\it Herschel} temperature map, the sites, G35.02N and G35.02S are seen with 
the warmer gas (T$_{d}$ $\sim$21-35 K) (see Figure~\ref{fig4}a), while the {\it Spitzer} dark clouds are traced in a 
temperature range of about 14--18~K. The horseshoe-like structure is also evident in the {\it Herschel} column density map and 
several condensations are seen toward this structure.

In the column density map, we employed the {\it clumpfind} algorithm to identify the clumps and their total column densities. 
Fourteen clumps are detected above a three sigma noise level in the MCG35.2. 
These clumps are labeled in Figure~\ref{fig4}b and their boundaries are also shown in Figure~\ref{fig4}c.
The mass of each clump is estimated using its total column density and can be determined using the equation:
\begin{equation}
M_{clump} = \mu_{H_2} m_H Area_{pix} \Sigma N(H_2)
\end{equation}
where $\mu_{H_2}$ is assumed to be 2.8, $Area_{pix}$ is the area subtended by one pixel, and 
$\Sigma N(\mathrm H_2)$ is the total column density. 
The mass of each {\it Herschel} clump is listed in Table~\ref{tab1}. 
The table also provides an effective radius and a peak column density of each clump. 
The clump masses vary between $\sim$110 M$_{\odot}$ and $\sim$4250 M$_{\odot}$. 
The peak column densities of these clumps also vary between $\sim$1.1 $\times$ 10$^{22}$ cm$^{-2}$ (A$_{V}$ $\sim$12 mag) and 
$\sim$14 $\times$ 10$^{22}$ cm$^{-2}$ (A$_{V}$ $\sim$150 mag) (see Table~\ref{tab1} and also Figure~\ref{fig4}b). 
There are at least ten massive clumps (M$_{clump}$ $\sim$500 -- 4250 M$_{\odot}$) distributed toward the horseshoe-like structure. 
\subsection{Kinematics of molecular gas}
\label{sec:coem} 
\begin{figure*}
\epsscale{1}
\plotone{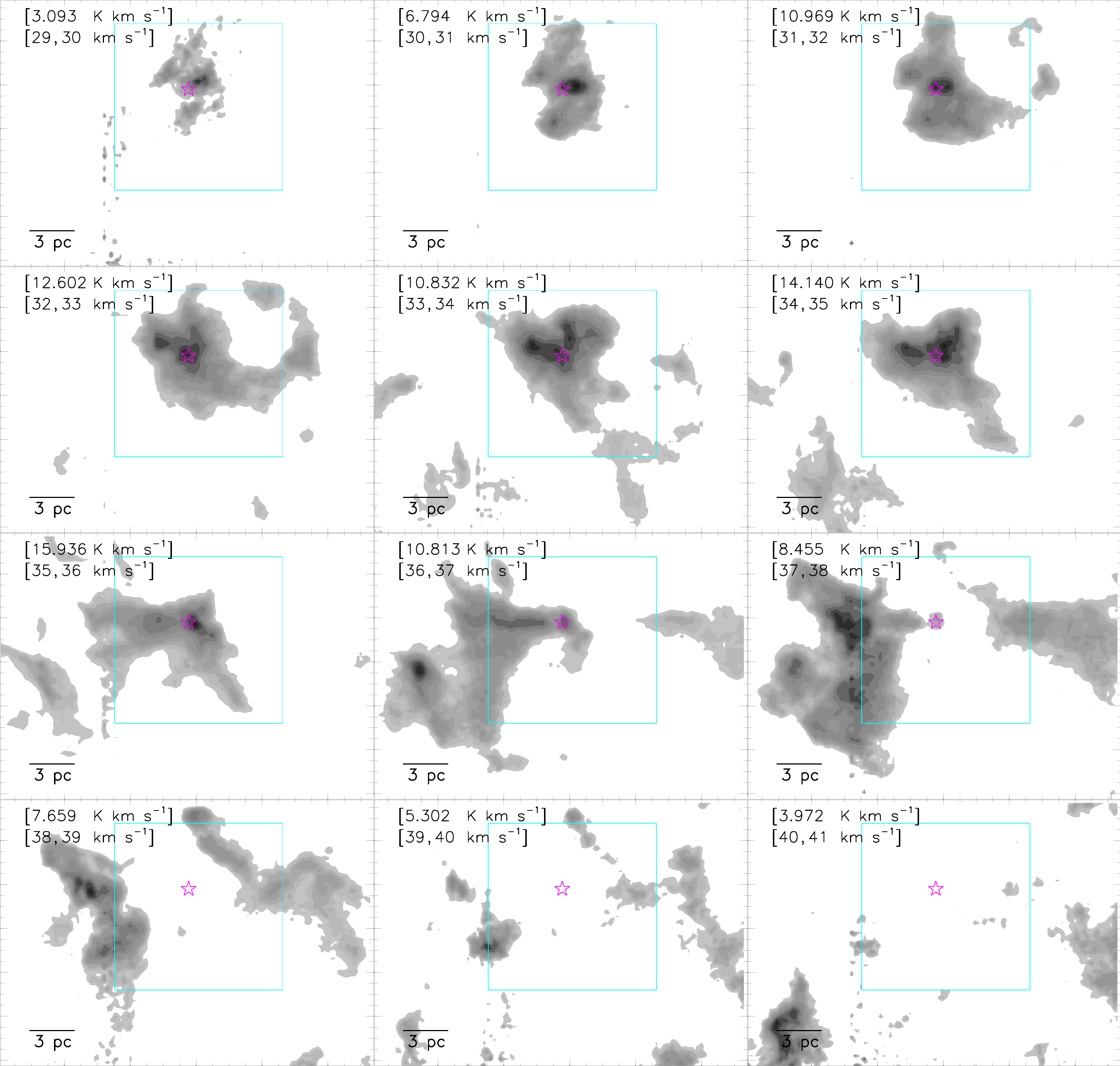}
\caption{\scriptsize Velocity channel contour maps of $^{13}$CO(J =1$-$0) emission. 
The molecular emission is integrated over a velocity interval, which is shown in each panel (in km s$^{-1}$). 
The contour levels are 20, 25, 30, 35, 40, 50, 60, 70, 80, 90, and 98\% of the peak value (in K km s$^{-1}$), 
which is also labeled in each panel. In each panel, the position of the NVSS 1.4 GHz peak is marked by a star symbol. 
In each panel, a solid box (in cyan) depicts the region seen at the intersection of the red-shifted and blue-shifted molecular components in the MCG35.2. 
The distribution of molecular gas also traces a possible complementary pair at 32-33 km s$^{-1}$ and 37-38 km s$^{-1}$ in the extended MCG35.2 complex.}
\label{fig5}
\end{figure*}

A kinematic analysis of the molecular gas in the MCG35.2 is presented in this section.
As mentioned before, the $^{13}$CO profile traces the MCG35.2 in a velocity range of 30--40 km s$^{-1}$ (see Figure~\ref{fig1}). 
In Figure~\ref{fig5}, we present the integrated GRS $^{13}$CO (J=1$-$0) velocity channel 
maps (at intervals of 1 km s$^{-1}$), showing two molecular components along the line of sight. 
The channel maps show a molecular component associated with the sites, G35.20N and G35.20S in a velocity range of 30--36 km s$^{-1}$ 
(having a velocity peak at $\sim$33 km s$^{-1}$), while the other molecular component is seen at 35--40 km s$^{-1}$ 
(having a velocity peak at $\sim$37 km s$^{-1}$). Hence, we infer the red-shifted and blue-shifted molecular components in the MCG35.2. 
The sites, G35.20N and G35.20S appear to be seen at the intersection of these molecular components. 
\citet{paron10} also utilized the GRS $^{13}$CO data and presented the integrated GRS $^{13}$CO intensity map and velocity 
channel maps mainly toward the sites, G35.20N and G35.20S \citep[see Figures 3 and 4 in][]{paron10}. 
However, the position-velocity analysis of the MCG35.2 is still lacking. 

\begin{figure*}
\epsscale{1}
\plotone{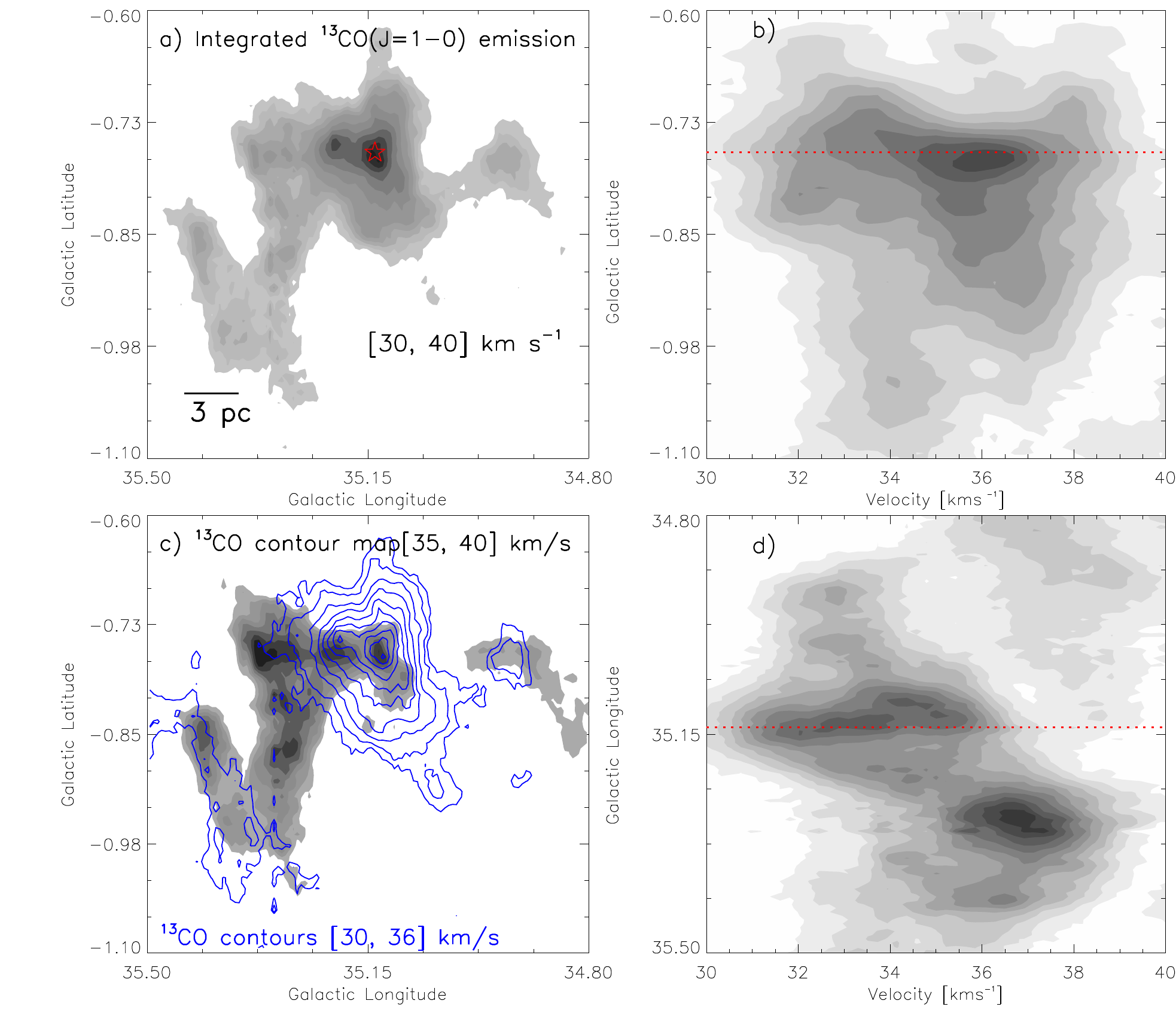}
\caption{\scriptsize  \textbf{Left panels; ``a" and ``c":} Integrated intensity maps of $^{13}$CO (J = 1-0) emission. 
The maps are similar to the one shown in Figure~\ref{fig1}. 
In top left panel ``a", the position of the NVSS 1.4 GHz peak is marked by a star. In bottom left panel ``c", the $^{13}$CO emission contours from 30 to 36 km s$^{-1}$ are overplotted on the 
$^{13}$CO emission map. 
The background $^{13}$CO emission map (from 35 to 40 km s$^{-1}$) is shown with levels of 
25.133 K km s$^{-1}$ $\times$ (0.35, 0.42, 0.5, 0.6, 0.7, 0.8, and 0.9). 
The $^{13}$CO contours (in blue) are shown with levels of 
55.165 K km s$^{-1}$ $\times$ (0.15, 0.2, 0.3, 0.4, 0.5, 0.6, 0.7, 0.8, 0.9, and 0.98).
\textbf{Right panels:} Position-velocity maps of $^{13}$CO (J = 1-0) emission. In top right panel ``b", Latitude-velocity distribution of $^{13}$CO. The CO emission is integrated over the longitude from 34$\degr$.8 to 35$\degr$.5.
In bottom right panel ``d", Longitude-velocity distribution of $^{13}$CO. The CO emission is integrated over the latitude from $-$0.$\degr$6 to $-$1.$\degr$1. 
In the right panels (i.e., position-velocity maps), a dotted red line shows the position of the NVSS 1.4 GHz peak. 
In right bottom panel ``d", the position-velocity map traces two peaks (a red-shifted and a blue-shifted) which are separated by 
 lower intensity intermediated velocity emission (i.e., a broad bridge feature; also see the text). 
 The detection of a broad bridge feature in the position-velocity map may indicate the signature of collisions between molecular components in the MCG35.2.}
\label{fig6}
\end{figure*}

In Figure~\ref{fig6}, we show the integrated $^{13}$CO intensity map and the position-velocity maps.  
In Figure~\ref{fig6}a, the integrated GRS $^{13}$CO intensity map is similar to that shown in Figure~\ref{fig1}.
The galactic position-velocity diagrams of the $^{13}$CO emission trace a noticeable velocity 
gradient (see Figures~\ref{fig6}b and~\ref{fig6}d). 
In Figure~\ref{fig6}c, the molecular emission integrated over 30 to 36 km s$^{-1}$ is also overlaid on the molecular map. 
In the background CO map, the molecular emission is shown from 35 to 40 km s$^{-1}$. 
Figure~\ref{fig6}c shows the spatial distribution of molecular gas linked with the red-shifted and blue-shifted molecular components in the MCG35.2. 
Figure~\ref{fig6}d also shows a red-shifted peak (at $\sim$37 km s$^{-1}$) and a blue-shifted peak (at $\sim$33 km s$^{-1}$) 
that are interconnected by lower intensity intermediated velocity emission, indicating the presence of a broad bridge feature. 
The broad bridge feature in the position-velocity diagram has been reported as an evidence of collisions 
between molecular clouds \citep{haworth15a,haworth15b}. 
Furthermore, \citet{torii16} reported that the complementary distribution is an expected output of cloud-cloud collisions, 
and the bridge feature at the intermediate velocity range can be explained as the turbulent gas excited at the interface of the collision. 
In the extended MCG35.2 complex, a possible complementary pair is 32-33 km s$^{-1}$ and 37-38 km s$^{-1}$ (see Figure~\ref{fig5}). 
We discuss the implication of our findings in more detail in the discussion Section~\ref{sec:disc}.
\subsection{GPIPS H-band Polarization}
\label{subsec:pol}
\begin{figure*}
\epsscale{0.82}
\plotone{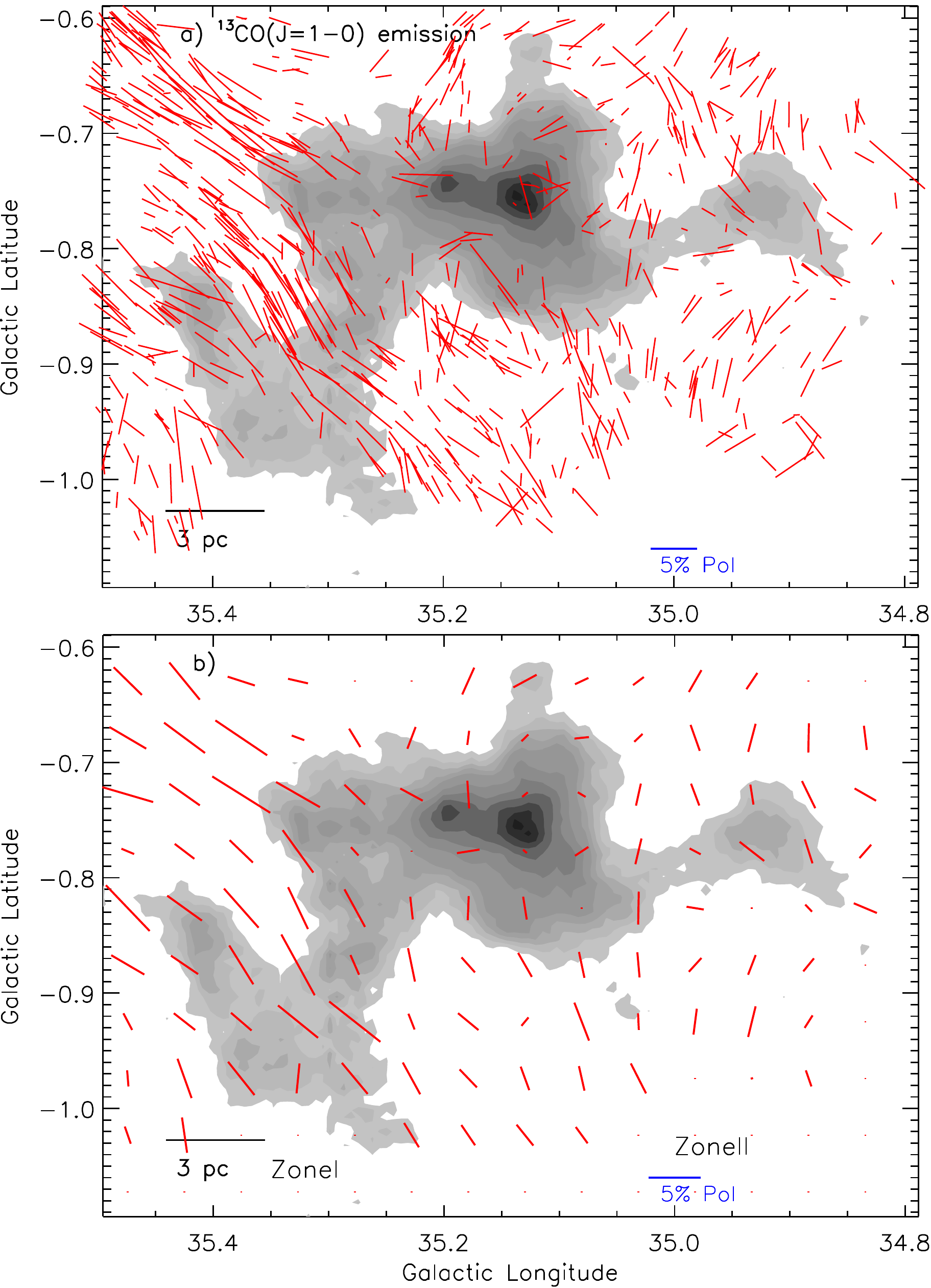}
\caption{\scriptsize The distribution of the GPIPS H-band polarization vectors and the GRS $^{13}$CO (J = 1-0) emission. 
The GPIPS H-band polarization vectors (in red) of 610 stars (with UF = 1 and $P/\sigma_p \ge$ 2) 
are overlaid on the $^{13}$CO map.
b) The mean polarization vectors (in red) are overplotted on the $^{13}$CO emission map. 
The mean polarization data are obtained by dividing the polarization spatial field into 14 $\times$ 10 equal parts 
and, a mean polarization value of H-band sources is computed within each specific part. 
In each panel, the degree of polarization is traced from the length of each vector. 
The orientations of the vectors show the galactic position angles of polarization in both the panels. 
A reference vector of 5\% is drawn in both the panels. 
In each panel, the background map is similar to the one shown in Figure~\ref{fig1}. 
In each panel, the scale bar at the bottom-left corner corresponds to 3 pc (at a distance of 2.0 kpc).}
\label{fig7}
\end{figure*}

In this section, we present the large scale morphology of the plane-of-the-sky projection of 
the magnetic field toward the MCG35.2. 
Using the GPIPS data, we study the polarization vectors of background stars that are selected 
based on the conditions (i.e. UF = 1 and $P/\sigma_p \ge$ 2). 
These conditions are very useful to select the reliable background stars and the confirmation of 
these background sources can also be obtained using the NIR color-color diagram \citep[e.g.][]{dewangan15}. 
As mentioned before, the polarization vectors of background stars give the field direction in the plane of the sky parallel to the direction of
polarization \citep{davis51}. 
In Figure~\ref{fig7}a, we display the H-band polarization vectors overlaid on the molecular map.  
To study the distribution of H-band polarization, we present mean polarization vectors in Figure~\ref{fig7}b. 
Our selected spatial filed is divided into 14 $\times$ 10 equal divisions 
and a mean polarization value is computed using the Q and U Stokes parameters of H-band sources 
located inside each division. 
In Figure~\ref{fig7}, the degree of polarization is traced by the length of a vector, whereas the angle of a vector indicates 
the polarization galactic position angle. 

The mean distribution of H-band vectors does not appear uniform. 
Our analysis reveals a noticeable change in the H-band starlight mean polarization angles between the ZoneI and ZoneII (see Figure~\ref{fig7}b).  
\subsection{Young stellar objects in the MCG35.2}
\label{subsec:phot1}
The knowledge of young populations helps to trace the ongoing star formation activity in a given embedded molecular cloud.
In this section, the young populations are investigated using their infrared excess for a wider field of view around the MCG35.2.
Various infrared photometric surveys (e.g., MIPSGAL, GLIMPSE, UKIDSS-GPS, and 2MASS) have been employed to 
explore the YSOs in the MCG35.2. 
In the following, a brief description of the selection of YSOs is provided.\\

\begin{figure*}
\epsscale{1}
\plotone{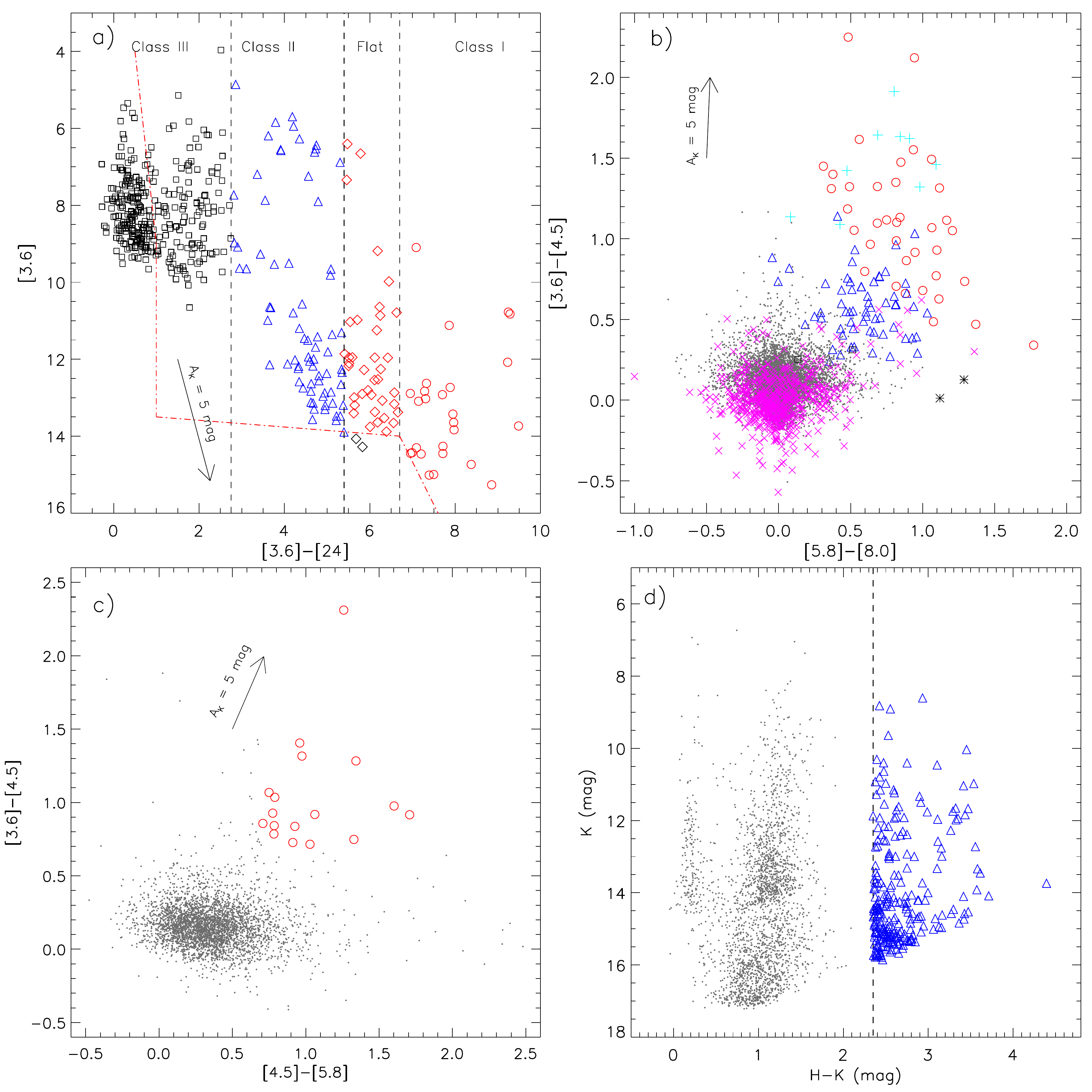}
\caption{\scriptsize Selection of YSOs in the star-forming complex, G35.20$-$0.74. 
a) GLIMPSE 3.6 $\mu$m, MIPSGAL 24 $\mu$m color-magnitude diagram of sources (i.e., GLIMPSE-MIPSGAL scheme; see the text for more details). 
The diagram depicts YSOs belonging to different evolutionary stages (see dashed lines). 
The zone of YSOs against contaminated candidates (galaxies and disk-less stars) is separated by dotted-dashed lines (in red) 
\citep[see][for more details]{rebull11}. Flat-spectrum and Class~III sources are shown by ``$\Diamond$'' and ``$\Box$'' symbols, respectively. 
b) GLIMPSE 3.6--8.0 $\mu$m color-color diagram of sources. 
The ``+", ``*'' and ``$\times$'' symbols show the  shocked emission, PAH-emitting galaxies and PAH-emission-contaminated apertures, 
respectively (see the text). Class~III sources are shown by ``$\Box$'' symbol. 
c) GLIMPSE 3.6--5.8 $\mu$m color-color diagram of sources. d) GPS-2MASS color-magnitude diagram (H$-$K/K) of the sources. 
In all the panels, we show Class~I (red circles) and Class~II (open blue triangles) YSOs. 
In the last three panels, the dots in gray color show the stars with only photospheric emissions. 
In the NIR H$-$K/K plot, we have plotted only 2503 out of 79775 stars with photospheric emissions. 
In the first three panels (i.e., color-color diagrams), an extinction vector for A$_{V}$ = 5 mag is shown \citep[using average extinction laws from][]{flaherty07}.}
\label{fig8}
\end{figure*}

1. In this scheme, we considered sources having detections in both the MIPSGAL 24 $\mu$m and GLIMPSE 3.6 $\mu$m bands (i.e. GLIMPSE-MIPSGAL scheme).
The GLIMPSE-MIPSGAL color-magnitude space ([3.6]$-$[24]/[3.6]) allows to depict the different stages of YSOs \citep{guieu10,rebull11,dewangan15}.
Following the conditions provided in \citet{guieu10}, the boundary of different stages of YSOs is marked in Figure~\ref{fig8}a. 
The color-magnitude space also exhibits the boundary of possible contaminants (i.e. galaxies and disk-less stars) \citep[see Figure~10 in][]{rebull11}.
In Figure~\ref{fig8}a, a total of 467 sources are presented in the color-magnitude space.
We find 137 YSOs (26 Class~I; 38 Flat-spectrum; 73 Class~II) and 328 Class~III sources. 
Additionally, two Flat-spectrum sources are seen in the boundary of possible contaminants and are not included in our selected young populations. 
Together, the selected YSO populations are free from the contaminants. \\
 
2. In this scheme, we considered sources having detections in all four GLIMPSE 3.6--8.0 $\mu$m bands (i.e. four GLIMPSE scheme). 
The GLIMPSE color-color space ([3.6]$-$[4.5] vs [5.8]$-$[8.0]) is used to identify the infrared excess sources. 
In the selected infrared excess sources, various possible contaminants (e.g. broad-line active galactic nuclei (AGNs), 
PAH-emitting galaxies, shocked emission blobs/knots, and PAH-emission-contaminated apertures) are removed using the \citet{gutermuth09} schemes.
Furthermore, the final selected YSOs are classified into different evolutionary stages based on their 
slopes of the GLIMPSE SED ($\alpha_{3.6-8.0}$) computed from 3.6 to 8.0 $\mu$m 
(i.e. Class~I ($\alpha_{3.6-8.0} > -0.3$), Class~II ($-0.3 > \alpha_{3.6-8.0} > -1.6$), 
and Class~III ($-1.6> \alpha_{3.6-8.0} > -2.56$)) \citep[e.g.,][]{lada06}. 
The GLIMPSE color-color space is presented in Figure~\ref{fig8}b. 
One can find more details about the YSO classifications based on the four GLIMPSE bands in \citet{dewangan11}. 
This scheme gives 99 YSOs (37 Class~I; 62 Class~II), 2 Class~III, and 589 contaminants.\\ 

3. In this scheme, we considered sources having detections in the first three GLIMPSE bands (except 8.0 $\mu$m band) (i.e. three GLIMPSE scheme). 
The color-color space ([4.5]$-$[5.8] vs [3.6]$-$[4.5]) is used to depict the infrared excess sources. 
We use color conditions, [4.5]$-$[5.8] $\ge$ 0.7 and [3.6]$-$[4.5] $\ge$ 0.7, to select protostars.
One can find more details about the YSO classifications based on the three GLIMPSE bands in \citet{hartmann05} and \citet{getman07}. 
This scheme yields 17 protostars in our selected region (see Figure~\ref{fig8}c). \\ 

4. In this scheme, we use the NIR color-magnitude space (H$-$K/K) to depict additional YSOs. 
We used a color H$-$K value (i.e. $\sim$2.35) that separates H$-$K excess sources from the rest of the population.
This color condition is chosen based on the color-magnitude plot of sources from the nearby control field 
(size of the selected region 12$\farcm$1 $\times$ 8$\farcm$7; centered at:  $l$ = 34$\degr$.898; $b$ = $-$1$\degr$.023).
Using this color H$-$K cut-off condition, 246 embedded YSOs are selected in the region probed in this work (see Figure~\ref{fig8}d).\\

Using the analysis of the photometric data at 1--24 $\mu$m, we obtain a total of 499 YSOs in our selected region around 
the MCG35.2 (as shown in Figure~\ref{fig1}). The positions of all these YSOs are shown in Figure~\ref{fig9}a. 
These YSOs are seen within the entire molecular cloud. 
\subsubsection{Study of distribution of YSOs}
\label{subsec:surfden}
\begin{figure*}
\epsscale{0.79}
\plotone{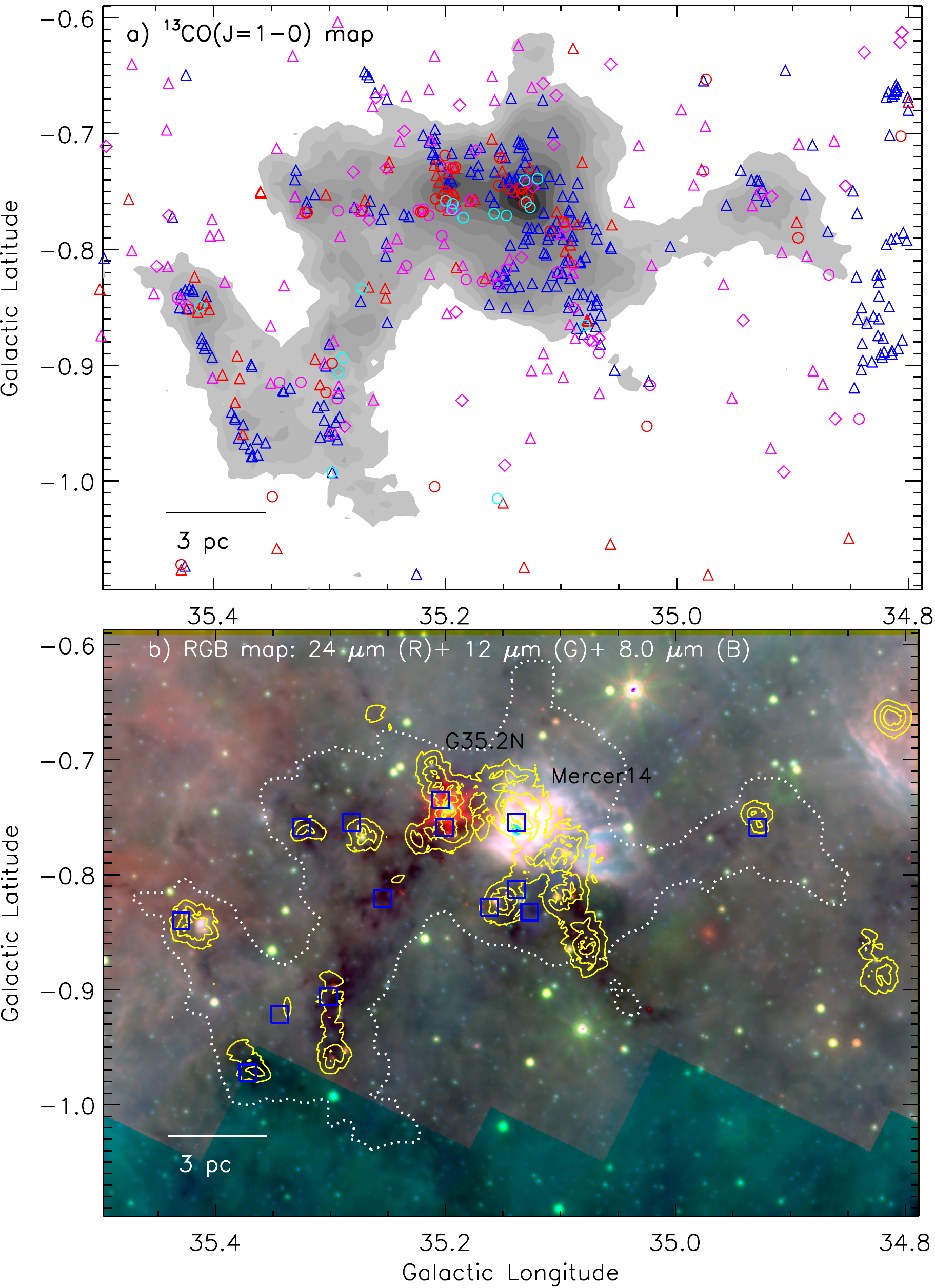}
\caption{\scriptsize  Spatial distribution of YSOs in the star-forming complex, G35.20$-$0.74. 
a) Based on the color-color and color-magnitude diagrams, the selected YSOs 
(Class~I (circles), Flat-spectrum (diamond), and Class~II (triangles)) are marked in the $^{13}$CO map. 
The background map is similar to the one shown in Figure~\ref{fig1}. 
The YSOs (in magenta) are identified using the GLIMPSE-MIPSGAL scheme (see Figure~\ref{fig8}a), 
while the YSOs selected using the other schemes (see the last three panels in Figure~\ref{fig8}) are marked in red (Class~I) and blue (Class~II) colors.
b) A {\it Spitzer}-{\it WISE} color-composite image is overlaid with surface density contours (in yellow) of all the identified YSOs. 
The surface density contours are shown at 5, 10, and 20 YSOs/pc$^{2}$, 
increasing from the outer to the inner regions. The {\it Herschel} clumps are also marked by squares (in blue) (see Figure~\ref{fig4}c). 
The CO emission is also overplotted by a dotted contour (in white) similar to those shown in Figure~\ref{fig2}a. 
Figures depict the distribution of clusters of YSOs in the MCG35.2.}
\label{fig9}
\end{figure*}

In this section, we study the spatial distribution of our selected YSOs using the standard surface density utility \citep[see][for more details]{gutermuth09,bressert10,dewangan11,dewangan15}, which can allow us to depict the individual groups or clusters of YSOs.
Using the nearest-neighbour (NN) technique, the surface density map of YSOs was obtained 
in a manner similar to that outlined in \citet{dewangan15} \citep[also see equation in][]{dewangan15}.  
The surface density map of all the selected 499 YSOs was constructed using 
a 5$\arcsec$ grid and 6 NN at a distance of 2.0 kpc. 
In Figure~\ref{fig9}b, the surface density contours of YSOs are presented and are drawn at 5, 10, 
and 20 YSOs/pc$^{2}$, increasing from the outer to the inner regions. 
The positions of {\it Herschel} clumps are also marked in Figure~\ref{fig9}b. 
We find noticeable YSOs clusters toward all the {\it Herschel} clumps in the MCG35.2. 
The {\it Herschel} clumps are very well correlated with the CO gas and YSOs. 
Furthermore, majority of the YSOs clusters are found toward the horseshoe-like structure.
The sites, G35.20N and G35.20S (including Ap2-1 and Mercer 14) are seen toward the {\it Herschel} clumps \#1--3, 
where the intense star formation activities are traced. Additionally, radio continuum emission is also detected in at least two out of these three clumps (see Figure~\ref{fig3}a).
\citet{krumholz08} suggested a threshold value of 1 gm cm$^{-2}$ (or corresponding column densities $\sim$3 $\times$ 10$^{23}$ cm$^{-2}$) 
for the birth of massive stars. Hence, theoretically, the birth of massive stars in these {\it Herschel} clumps is likely (see column densities in Table~\ref{tab1}). 
Previously, 
\citet{sanchez14} referred the G35.20N as the main site of MSF \citep[also see][]{beltran16}. 
In Section~\ref{sec:hh}, we find that the H\,{\sc ii} region associated with Ap2-1 is very young (i.e., $\sim$4100--32800 yr) and is excited by a radio spectral type of B0.5V star.
In addition to the YSOs clusters, there is also ongoing MSF in the {\it Herschel} clumps \#1--3 located at the base of the horseshoe-like structure. 
It is also worth mentioning that these three {\it Herschel} clumps have masses more than 1000 M$_{\odot}$ and 
peak column densities between $\sim$4.2 $\times$ 10$^{22}$ cm$^{-2}$ (A$_{V}$ $\sim$45 mag) and 
$\sim$14 $\times$ 10$^{22}$ cm$^{-2}$ (A$_{V}$ $\sim$150 mag) (see Table~\ref{tab1}). 
The radio continuum emission is not observed toward the remaining {\it Herschel} clumps (i.e. \#4-14), 
indicating the absence of the H\,{\sc ii} regions in these clumps. Hence, it implies that these remaining clumps 
appear to harbor the clusters of low-mass stars. 
The study of distribution of YSOs convincingly illustrates the star formation activities (including MSF) in the MCG35.2. 
\section{Discussion}
\label{sec:disc}
A recent demonstration of the tool of a broad-bridge feature concerning the cloud-cloud collision 
is given by the position velocity analysis of molecular gas. Recently, \citet{haworth15a} and \citet{haworth15b} carried 
out simulations in favor of this understanding \citep[also see][]{torii11,torii15,torii16,fukui14,fukui16,dewangan15,baug16}. 
In addition to the broad bridge feature, one can also expect the spatially complementary distribution between the two clouds 
in the sites of the cloud-cloud collision \citep[e.g.,][]{torii16}. 
\citet{fukui14} and \citet{torii15} discussed that the cloud-cloud collision process can also trigger the birth of massive stars \citep[also see][]{fukui16,torii16}.  \citet{fukui16} pointed out that the H$_{2}$ column density is the critical parameter in the cloud-cloud collision 
scenario and also reported a threshold value of 10$^{23}$ cm$^{-2}$ for the birth of the massive star clusters 
such as RCW38. They further suggested that a single O star can be formed at molecular column density 
of 10$^{22}$ cm$^{-2}$. 
There are some star-forming regions reported in the literature where the cloud-cloud collision process is likely and
\citet{torii16} provided a Table~2 containing the physical parameters of the cloud-cloud collisions 
estimated in the earlier studies. 

In Section~\ref{subsec:phot1}, the photometric analysis of point-like sources has revealed 
ongoing star formation activities throughout the MC35.20. However, 
the majority of the YSOs clusters are distributed toward the horseshoe-like structure where ten massive 
clumps are detected. The intense star formation activities are also investigated at the base of this structure where 
the previously known sites, G35.20N and G35.20S are located. 
Recently, \citet{beltran16} reported the presence of a binary system of ultra-compact/hyper-compact H\,{\sc ii} 
regions at the geometrical center of the radio jet in G35.20N. They further found that the binary system, which is 
associated with a Keplerian rotating disk, contains two B-type stars of 11 and 6 M$_{\odot}$. 
The sites, G35.20N and Ap2-1 are revealed as the active sites of MSF and are extended by about 4.5 pc.
Furthermore, the sites, G35.20N and G35.20S (including Ap2-1 and Mercer 14) are found at the intersection of two molecular components 
(having peaks at 33 and 37 km s$^{-1}$) in the MC35.20 (see Section~\ref{sec:coem}). 
The velocity separation of the two cloud components is depicted to be $\sim$4 km s$^{-1}$. 
We also investigate a broad bridge feature in the velocity space and the complementary distribution of 
the two molecular components in the velocity channel maps. 
All these characteristic features of cloud-cloud collision are evident in the MCG35.2, 
favouring the onset of the cloud-cloud collision process. Using the physical separation (i.e. $\sim$4.5 pc) 
and the velocity separation (i.e. $\sim$4 km s$^{-1}$) of the two colliding cloud components in the MCG35.2, 
we estimate the typical collision timescale to be $\sim$1.1 Myr. 
The dynamical age of the H\,{\sc ii} region associated with Ap2-1 is found to be $\sim$4100--32800 yr. 
An average age of the Class~I and Class~II YSOs is reported to be $\sim$0.44 Myr and $\sim$1--3 Myr, respectively \citep{evans09}. 
Considering these timescales, it appears that the birth of massive stars and the youngest populations 
is influenced by the cloud-cloud collision. The observed high column densities toward the clumps 
further support this interpretation (see Table~\ref{tab1}), which contain massive stars. 
However, the stellar populations older than the collision timescale may have formed prior to the collision. 
Hence, we cannot also rule out the star formation before the collision in the MC35.20.

Another result presented in this study is that a noticeable change in the distribution of the starlight H-band polarization vectors in the MC35.20.  
It is known that the large scale magnetic field structure projected in the sky plane can be inferred using the starlight H-band polarization data and 
the aligned dust is utilized to probe magnetic fields. Taken this fact, the interaction of molecular components could influence the dust distribution 
and appears to be responsible for the variation in the polarization angles. 
\section{Summary and Conclusions}
\label{sec:conc}
In this paper, we have utilized the multi-wavelength data covering from optical-H$\alpha$, NIR, and radio wavelengths and have carried out 
an extensive study of the MCG35.2. In the following, the important outcomes of this work are presented.\\
$\bullet$ The molecular gas emission in the direction of the MCG35.2 is traced in a velocity range of 30--40 km s$^{-1}$. There are two noticeable molecular components (having velocity peaks at 33 and 37 km s$^{-1}$) detected in the MCG35.2.\\
$\bullet$ The most prominent structure in the MCG35.2 is the horseshoe-like morphology traced in the infrared and millimeter images.\\ 
$\bullet$ The distribution of ionized emission toward Ap2-1 observed in the MAGPIS 20 cm continuum map and H$\alpha$ image is 
almost spherical. The ionizing photon flux value estimated at 20 cm corresponds to a single ionizing star of B0.5V--B0V spectral type.\\
$\bullet$ The position-velocity analysis of $^{13}$CO emission depicts a broad bridge feature which is defined as a structure 
having two velocity peaks (a red-shifted (at 37 km s$^{-1}$) and a blue-shifted (at 33 km s$^{-1}$)) interconnected by lower intensity intermediated velocity emission. 
Such feature can indicate the signature of a collision between molecular components in the MCG35.2. 
These two velocity peaks are separated by $\sim$4 km s$^{-1}$. 
The study of molecular line data also shows a possible complementary pair at 32-33 km s$^{-1}$ and 37-38 km s$^{-1}$ within the extended MCG35.2 complex.\\ 
$\bullet$ Fourteen {\it Herschel} clumps are identified in the MCG35.2 and ten out of these fourteen clumps are seen toward the horseshoe-like structure.\\  
$\bullet$ The analysis of photometric data at 1--24 $\mu$m reveals a total of 499 YSOs and these YSOs are found in the clusters 
distributed within the molecular cloud. Interestingly, majority of the clusters of YSOs are found toward the horseshoe-like structure. 
At the base of the horseshoe-like structure, very intense star formation activities (including massive stars) are found.\\
$\bullet$ The analysis of the GPIPS data reveals a variation in the distribution of the polarization position angles of background starlight in the MCG35.2.\\

Based on all our observational findings, we conclude that the star formation activities in the MCG35.2 are influenced by the cloud-cloud collision process.
\acknowledgments
We thank the anonymous reviewer for constructive comments and suggestions, which greatly improved the scientific contents of the paper. 
We are grateful to Dr. A. Luna for proving the IDL-based program for the analysis of CO line data. 
The research work at Physical Research Laboratory is funded by the Department of Space, Government of India. 
This work is based on data obtained as part of the UKIRT Infrared Deep Sky Survey. This publication 
made use of data products from the Two Micron All Sky Survey (a joint project of the University of Massachusetts and 
the Infrared Processing and Analysis Center / California Institute of Technology, funded by NASA and NSF), archival 
data obtained with the {\it Spitzer} Space Telescope (operated by the Jet Propulsion Laboratory, California Institute 
of Technology under a contract with NASA). 
This publication makes use of molecular line data from the Boston University-FCRAO Galactic
Ring Survey (GRS). The GRS is a joint project of Boston University and Five College Radio Astronomy Observatory, 
funded by the National Science Foundation (NSF) under grants AST-9800334, AST-0098562, and AST-0100793.  
This publication makes use of the Galactic Plane Infrared Polarization Survey (GPIPS). 
The GPIPS was conducted using the {\it Mimir} instrument, jointly developed at Boston University and Lowell Observatory
and supported by NASA, NSF, and the W.M. Keck Foundation. 
This paper makes use of data obtained as part of the INT Photometric H$\alpha$ Survey of the Northern Galactic 
Plane (IPHAS, www.iphas.org) carried out at the Isaac Newton Telescope (INT). The INT is operated on the 
island of La Palma by the Isaac Newton Group in the Spanish Observatorio del Roque de los Muchachos of 
the Instituto de Astrofisica de Canarias. The IPHAS data are processed by the Cambridge Astronomical Survey 
Unit, at the Institute of Astronomy in Cambridge.

 \begin{table*}
  \tiny
\setlength{\tabcolsep}{0.05in}
\centering
\caption{List of surveys (covering from radio, NIR, and optical-H$\alpha$ wavelengths) utilized in this work}
\label{ftab1}
\begin{tabular}{lcccr}
\hline 
  Survey  &  Wavelength(s)       &  Resolution ($\arcsec$)        &  Reference  \\   
\hline
\hline 
 NRAO VLA Sky Survey (NVSS)                                                             &21 cm            & $\sim$45        &       \citet{condon98}    \\     
 Multi-Array Galactic Plane Imaging Survey (MAGPIS)                             & 20 cm                       & $\sim$6          & \citet{helfand06}\\
 Galactic Ring Survey (GRS)                                                                   & 2.7 mm; $^{13}$CO (J = 1--0) & $\sim$45        &\citet{jackson06}\\
APEX Telescope Large Area Survey of the Galaxy (ATLASGAL)                 &870 $\mu$m                     & $\sim$19.2        &\citet{schuller09}\\
{\it Herschel} Infrared Galactic Plane Survey (Hi-GAL)                              &70, 160, 250, 350, 500 $\mu$m                     & $\sim$5.8, $\sim$12, $\sim$18, $\sim$25, $\sim$37         &\citet{molinari10}\\
{\it Spitzer} MIPS Inner Galactic Plane Survey (MIPSGAL)                                         &24 $\mu$m                     & $\sim$6         &\citet{carey05}\\ 
Wide Field Infrared Survey Explorer (WISE)                                             &3.4, 4.6, 12, 22 $\mu$m                   & $\sim$6, $\sim$6.5, $\sim$6.5, $\sim$12           &\citet{wright10}\\ 
{\it Spitzer} Galactic Legacy Infrared Mid-Plane Survey Extraordinaire (GLIMPSE)       &3.6, 4.5, 5.8, 8.0  $\mu$m                   & $\sim$2, $\sim$2, $\sim$2, $\sim$2           &\citet{benjamin03}\\
UKIRT Wide-field Infrared Survey for H2 (UWISH2)                                 &2.12 $\mu$m                   & $\sim$0.8          &\citet{froebrich11a}\\ 
UKIRT NIR Galactic Plane Survey (GPS)                                                 &1.25--2.2 $\mu$m                   &$\sim$0.8           &\citet{lawrence07}\\ 
Galactic Plane Infrared Polarization Survey (GPIPS)                                 &1.6 $\mu$m                   & $\sim$1.5          &\citet{clemens12}\\
Two Micron All Sky Survey (2MASS)                                                 &1.25--2.2 $\mu$m                  & $\sim$2.5          &\citet{skrutskie06}\\
Isaac Newton Telescope Photometric H$\alpha$ Survey of the Northern Galactic Plane (IPHAS) &0.6563 $\mu$m                  & $\sim$1          &\citet{drew05}\\
\hline          
\end{tabular}
\end{table*}

\begin{table*}
\setlength{\tabcolsep}{0.05in}
\centering
\caption{The properties of the {\it Herschel} clumps identified in the star-forming complex, G35.20$-$0.74 
(see Figures~\ref{fig4}b and~\ref{fig4}c). Column~1 lists the IDs assigned to the clump. Table also provides 
positions, deconvolved effective radius (R$_{c}$), clump mass (M$_{clump}$), peak column density (N(H$_{2}$)), and extinction ($A_V=1.07 \times 10^{-21}~N(\mathrm H_2)$).}
\label{tab1}
\begin{tabular}{lcccccr}
\hline 
ID & {\it l}       & {\it b}     & R$_{c}$ & M$_{clump}$ & peak N(H$_{2}$) & A$_{V}$\\
    &  [degree] & [degree] &  (pc)      &($M_\odot$)   &(10$^{22}$ cm$^{-2}$) &(mag) \\
\hline
\hline
       1  &	 35.138     &	-0.755   &	 1.5	&    4250 & 14     &     150  \\
       2  &	 35.201     &	-0.759   &	 1.2	&    3100 & 8.5    &       91  \\
       3  &	 35.205     &	-0.735   &	 0.9	&    1030 & 4.2    &       45  \\
       4  &	 35.302     &	-0.906   &	 1.0	&     900 & 2.9    &       31  \\
       5  &	 35.162     &	-0.829   &	 0.9	&     700 & 2.4    &       26  \\
       6  &	 35.255     &	-0.821   &	 1.0	&     960 & 1.9    &       20  \\
       7  &	 35.325     &	-0.759   &	 0.8	&     540 & 1.8    &       19  \\
       8  &	 35.282     &	-0.755   &	 0.9	&     700 & 1.5    &       16  \\
       9  &	 35.345     &	-0.922   &	 0.4	&     110 & 1.4    &       15  \\
      10  &	 35.138     &	-0.813   &	 0.9	&     670 & 1.4    &       15  \\
      11  &	 34.928     &	-0.759   &	 0.9	&     550 & 1.3    &       14  \\
      12  &	 35.372     &	-0.972   &	 0.5	&     160 & 1.2    &       13  \\
      13  &	 35.127     &	-0.832   &	 0.8	&     500 & 1.2    &       13  \\
      14  &	 35.430     &	-0.840   &	 0.4	&     140 & 1.1    &       12  \\
\hline          
\end{tabular}
\end{table*}


\begin{thebibliography}{}
%
\bibitem[Baug et al.(2016)]{baug16}
Baug, T., Dewangan, L.~K., Ojha, D.~K., et al. 2016, ApJ, arXiv:1609.06440

\bibitem[Beltr\'{a}n et al.(2016)]{beltran16} 
Beltr\'{a}n, M.~T., Cesaroni, R., Moscadelli, L., et al. 2016, A\&A, 593, 49

\bibitem[Benjamin et al.(2003)]{benjamin03}
Benjamin, R.~A.,Churchwell, E., Babler, B.~L., et al. 2003, PASP, 115, 953

\bibitem[Birks et al.(2006)]{birks06} 
Birks, J.~R., Fuller, G.~A., \& Gibb, A.~G. 2006, A\&A, 458, 181

\bibitem[Bisbas et al.(2009)]{bisbas09}
Bisbas, T.~G., W\"{u}nsch, R., Whitworth, A.~P., \& Hubber, D.~A. 2009, A\&A, 497, 649

\bibitem[Bohlin et al.(1978)]{bohlin78}
Bohlin, R.~C., Savage, B.~D., \& Drake, J.~F. 1978, ApJ, 224, 13233

\bibitem[Bressert et al.(2010)]{bressert10}
Bressert, E., Bastian, N., Gutermuth, R., et al. 2010, MNRAS, 409, 54

\bibitem[Carey et al.(2005)]{carey05}
Carey, S. J., Noriega-Crespo, A., Price, S.~D., et al. 2005, BAAS, 37, 1252

\bibitem[Casali et al.(2007)]{casali07}
Casali, M., Adamson, A., Alves de Oliveira, C., et al. 2007, A\&A, 467, 777

\bibitem[Clemens et al.(2012)]{clemens12}
Clemens, D.~P., Pavel, M.~D., \& Cashman, L.~R. 2012, ApJS, 200, 21

\bibitem[Condon et al.(1998)]{condon98}
Condon, J.~J., Cotton, W.~D., Greisen, E.~W., et al. 1998, AJ, 115, 1693

\bibitem[Cyganowski et al.(2008)]{cyganowski08} 
Cyganowski, C.~J., Whitney, B.~A., Holden, E., et al. 2008, AJ, 136, 2391

\bibitem[Davis \& Greenstein(1951)]{davis51}
Davis, L., Jr., \& Greenstein, J. L. 1951, ApJ, 114, 206

\bibitem[Dewangan \& Anandarao(2011)]{dewangan11}
Dewangan, L.~K., \& Anandarao, B.~G 2011, MNRAS, 414, 1526

\bibitem[Dewangan et al.(2015)]{dewangan15}
Dewangan, L.~K., Luna, A., Ojha, D.~K., et al.  2015, ApJ, 811, 79

\bibitem[Dewangan et al.(2016)]{dewangan16}
Dewangan, L.~K., Ojha, D.~K., Luna, A., et al.  2016, ApJ, 819, 66

\bibitem[Drew et al.(2005)]{drew05}
Drew, J.~E., Greimel, R., Irwin, M.J., et al. 2005, MNRAS, 362, 753

\bibitem[Dyson \& Williams(1980)]{dyson80}
Dyson, J.~E., \& Williams, D.~A. 1980, Physics of the Interstellar Medium (New York: Halsted Press)

\bibitem[Evans et al.(2009)]{evans09}
Evans, N.~J., II, Dunham, M.~M., J\o{}rgensen, J.~K., et al. 2009, ApJS, 181, 321

\bibitem[Flaherty et al.(2007)]{flaherty07}
Flaherty, K.~M., Pipher, J.~L., Megeath, S.~T., et al. 2007, ApJ, 663, 1069

\bibitem[Froebrich et al.(2011a)]{froebrich11a} 
Froebrich, D., Davis, C.~J., Ioannidis G., et al. 2011a, MNRAS, 413, 480

\bibitem[Froebrich \& Ioannidis(2011b)]{froebrich11} 
Froebrich,D. \& Ioannidis, G. 2011b, MNRAS, 418, 1375

\bibitem[Fukui et al.(2014)]{fukui14} 
Fukui, Y., Ohama, A., Hanaoka, N., et al.\ 2014, ApJ, 780, 36 

\bibitem[Fukui et al.(2016)]{fukui16} 
Fukui, Y., Torii, K., Ohama, A., et al. 2016, ApJ, 820, 26 

\bibitem[Furukawa et al.(2009)]{furukawa09} 
Furukawa, N., Dawson, J.~R., Ohama, A., et al.\ 2009, ApJL, 696, L115 

\bibitem[Getman et al.(2007)]{getman07} 
Getman, K.~V., Feigelson, E.~D., Garmire,G., Broos, P., \& Wang, J. 2007, ApJ, 654, 316 

\bibitem[Griffin et al.(2010)]{griffin10} 
Griffin, M.~J., Abergel, A., Abreu, A, et al. 2010, A\&A, 518L, 3

\bibitem[Guieu et al.(2010)]{guieu10}
Guieu, S., Rebull, L.~M., Stauffer, J.~R., et al. 2010, ApJ, 720, 46

\bibitem[Gutermuth et al.(2009)]{gutermuth09}
Gutermuth, R.~A., Megeath, S.~T., Myers, P.~C., et al. 2009, ApJS, 184, 18

\bibitem[Gutermuth \& Heyer(2015)]{gutermuth15}
Gutermuth, R.~A., \& Heyer,M. 2015, AJ, 149, 64

\bibitem[Habe \& Ohta(1992)]{habe92} 
Habe, A., \& Ohta, K.\ 1992, PASJ, 44, 203 

\bibitem[Hartmann et al.(2005)]{hartmann05} 
Hartmann, L., Megeath, S.~T., Allen, L., et al. 2005, ApJ, 629, 881

\bibitem[Haworth et al.(2015a)]{haworth15a} 
Haworth, T.~J., Tasker, E.~J., Fukui, Y., et al. 2015a, MNRAS, 450, 10

\bibitem[Haworth et al.(2015b)]{haworth15b} 
Haworth, T.~J., Shima, K., Tasker, E.~J., et al. 2015b, MNRAS, 454, 1634

\bibitem[Helfand et al.(2006)]{helfand06}
Helfand, D.~J., Becker, R.~H., White, R.~L., Fallon, A., \& Tuttle, S. 2006, AJ, 131, 2525 

\bibitem[Hildebrand(1983)]{hildebrand83} 
Hildebrand, R.~H. 1983, QJRAS, 24, 267

\bibitem[Inoue \& Fukui(2013)]{inoue13} 
Inoue, T., \& Fukui, Y. 2013, ApJL, 774, 31

\bibitem[Jackson et al.(2006)]{jackson06} 
Jackson, J.~M., Rathborne, J.~M., Shah, R.~Y., et al. 2006, ApJS, 163, 145

\bibitem[Kauffmann et al.(2008)]{kauffmann08}
Kauffmann, J., Bertoldi, F., Bourke, T.~L., Evans, II, N.~J.,\&  Lee, C.~W. 2008, ApJ, 487, 993

\bibitem[Krumholz \& McKee(2008)]{krumholz08}
Krumholz, M.,~R., \& McKee, C.,~F. 2008, Nature, 451, 1082

\bibitem[Kwan(1997)]{kwan97} 
Kwan, J. 1997, ApJ, 489, 284

\bibitem[Lada et al.(2006)]{lada06}
Lada, C.~J., Muench, A.~A., Luhman, K.~L., et al. 2006, AJ, 131, 1574

\bibitem[Lawrence et al.(2007)]{lawrence07}
Lawrence, A., Warren, S.~J., Almaini, O., et al. 2007, MNRAS, 379, 1599

\bibitem[Mallick et al.(2015)]{mallick15}
Mallick, K.~K., Ojha, D.~K., Tamura, M., et al. 2015, MNRAS, 447, 2307

\bibitem[Matsakis et al.(1976)]{matsakis76}
Matsakis, D.~N., Evans, N.~J., II, Sato, T., \& Zuckerman, B. 1976, AJ, 81, 172

\bibitem[Molinari et al.(2010)]{molinari10}
Molinari, S., Swinyard, B., Bally, J., et al., 2010, A\&A, 518, L100

\bibitem[Mooney et al.(1995)]{mooney95} 
Mooney, T., Sievers, A., \& Mezger, P.~G., et al. 1995, A\&A, 299, 869

\bibitem[Ott(2010)]{ott10}
Ott, S. 2010, in Astronomical Society of the Pacic Conference
Series, Vol. 434, Astronomical Data Analysis Software and
Systems XIX, ed. Y. Mizumoto, K.-I. Morita, \& M. Ohishi, 139

\bibitem[Ohama et al.(2010)]{ohama10} 
Ohama, A., Dawson, J.~R., Furukawa, N., et al.\ 2010, ApJ, 709, 975 

\bibitem[Panagia(1973)]{panagia73} 
Panagia, N. 1973, AJ, 78, 929 

\bibitem[Paron \& Weidmann(2010)]{paron10} 
Paron,S., \& Weidmann, W. 2010, MNRAS, 408, 2487

\bibitem[Poglitsch et al.(2010)]{poglitsch10}	
Poglitsch, A., Waelkens, C., Geis, N., et al. 2010, A\&A, 518L, 2

\bibitem[Peretto \& Fuller(2009)]{peretto09}
Peretto, N., \& Fuller, G.~A. 2009, A\&A, 505, 405

\bibitem[Qiu et al.(2013)]{qiu13} 
Qiu, K., Zhang, Q., Menten, K.~M., Liu, H.~B., \& Tang, Y.~W. 2013, ApJ, 779, 182 

\bibitem[Rebull et al.(2011)]{rebull11}
Rebull, L.~M., Guieu, S., Stauffer, J.~R., et al. 2011, ApJS, 193, 25

\bibitem[S\'{a}nchez et al.(2014)]{sanchez14} 
S\'{a}nchez-Monge, \'{A}., Beltr\'{a}n, M.~T., Cesaroni, R., et al. 2014, A\&A, 569, 11

\bibitem[Schuller et al.(2009)]{schuller09}
Schuller, F., Menten, K.~M., Contreras, Y., et al. 2009, A\&A, 504, 415

\bibitem[Seta et al.(1998)]{seta98} 
Seta, M., Hasegawa, T., Dame, T., et al. 1998, ApJ, 505, 286

\bibitem[Skrutskie et al.(2006)]{skrutskie06}
Skrutskie, M.~F., Cutri, R.~M., Stiening, R., et al. 2006, AJ, 131, 1163

\bibitem[Tan et al.(2014)]{tan14} 	
Tan, J.~C., Beltr\'an, M.~T., Caselli, P., et al. 2014, in Protostars and Planets VI, ed. H. Beuther et al. (Tucson, AZ: Univ. Arizona Press), 149

\bibitem[Torii et al.(2011)]{torii11} 
Torii, K., Enokiya, R., Sano, H., et al. 2011, ApJ, 738, 46 

\bibitem[Torii et al.(2015)]{torii15} 
Torii, K., Hasegawa, K., Hattori, Y., et al. 2015, ApJ, 806, 7 

\bibitem[Torii et al.(2016)]{torii16} 
Torii, K., Hattori, Y., Hasegawa, K., et al. 2016, arXiv:1612.09458

\bibitem[Williams et al.(1994)]{williams94} 
Williams, J. P., de Geus, E. J., \& Blitz, L. 1994, ApJ, 428, 693

\bibitem[Wright et al.(2010)]{wright10}
Wright, E.~L., Eisenhardt, P.~R.~M., Mainzer, et al. 2010, AJ, 140, 1868

\bibitem[Zhang et al.(2009)]{zhang09} 
Zhang, B., Zheng, X.~W., Reid, M.~J., Menten, K.~M., et al. 2009, ApJ, 693, 419

\bibitem[Zinnecker \& Yorke(2007)]{zin07} 
Zinnecker, H., \& Yorke, H.~W. 2007, ARA\&A, 45, 481 

\end{thebibliography}
\end{document}